\documentstyle[11pt,amssymb]{article}                                     

\def\AFOUR{%
\setlength{\textheight}{9.0in}%
\setlength{\textwidth}{5.75in}%
\setlength{\topmargin}{-0.375in}%
\hoffset=-.5in%
\renewcommand{\baselinestretch}{1.17}%
\setlength{\parskip}{6pt plus 2pt}%
}
\AFOUR                                           
\def\car{\mathop{\square}}
\def\carre#1#2{\raise 2pt\hbox{$\scriptstyle #1$}\car_{#2}}

\makeatletter
\def\section{\@startsection {section}{1}{\z@}{-3.5ex plus -1ex minus
 -.2ex}{2.3ex plus .2ex}{\large\bf}}
\def\subsection{\@startsection{subsection}{2}{\z@}{-3.25ex plus -1ex minus
 -.2ex}{1.5ex plus .2ex}{\normalsize\bf}}
\makeatother
\makeatletter
\@addtoreset{equation}{section}

\makeatother
\newcommand{\nc}{\newcommand}
\newcommand{\rnc}{\renewcommand}
\nc{\be}{\begin{equation}}
\nc{\ee}{\end{equation}}
\nc{\bea}{\begin{eqnarray}}
\nc{\eea}{\end{eqnarray}}

\def\slash#1{\setbox0=\hbox{$#1$}#1\hskip-\wd0\hbox to\wd0{\hss\sl/\/\hss}}

\def\href#1#2{{#2}}

\rnc{\a}{\alpha}
\nc{\ab}{\bar{\a}}
\nc{\ap}{\a^{+}}
\nc{\abm}{\ab^{-}}
\rnc{\b}{\beta}
\nc{\bb}{\bar{\b}}
\nc{\bbp}{\bb_{\zb}^{+}}
\nc{\bm}{\b_{z}^{-}}
\nc{\oa}{\overline{\a}}
\nc{\ob}{\overline{\b}}
\rnc{\gg}{\gamma}
\rnc{\d}{\delta}
\nc{\f}{\phi}
\nc{\fb}{\bar{\phi}}
\nc{\vf}{\varphi}
\nc{\p}{\psi}

\rnc{\c}{\chi}
\nc{\la}{\lambda}
\nc{\m}{{\mathrm m}}
\nc{\n}{\nu}
\rnc{\o}{\omega}
\nc{\Om}{\Omega}
\rnc{\t}{\theta}
\nc{\eps}{\epsilon}
\rnc{\S}{\Sigma}
\nc{\F}{\Phi}
\nc{\trac}[2]{{\textstyle\frac{#1}{#2}}}
\nc{\ex}[1]{\mbox{e}^{\,\textstyle#1}}
\nc{\mat}[4]{\left(\begin{array}{cc}#1&#2\\#3&#4\end{array}\right)}
\nc{\som}[9]{\left(\begin{array}{ccc}#1&#2&#3\\#4&#5&#6\\#7&#8&#9%
\end{array}\right)}
\nc{\tr}{\mathop{\mbox{tr}}\nolimits}
\nc{\ad}{\mathop{\mbox{ad}}\nolimits}
\nc{\Tr}{\mathop{\mbox{Tr}}\nolimits}
\nc{\Det}{\mathop{\mbox{Det}}\nolimits}
\nc{\rk}{\mathop{\mbox{rk}}\nolimits}
\nc{\ra}{\rightarrow}
\nc{\Ra}{\Rightarrow}
\nc{\LRa}{\Leftrightarrow}
\nc{\ot}{\otimes}
\rnc{\ss}{\subset}
\nc{\nul}{\noindent\underline}
\nc{\non}{\nonumber\\}
\nc{\subs}[1]{{\vspace*{0.5cm}}%
{\noindent\underline{#1}}{\addcontentsline{toc}{subsection}{#1}}%
{\vspace*{0.3cm}}}
\nc{\zb}{\bar{z}}
\rnc{\lg}{\frak{g}}
\nc{\lt}{\frak{t}}
\nc{\lk}{\frak{k}}
\nc{\lh}{\frak{h}}
\nc{\pik}{\Pi_{\lk}}
\nc{\pip}{\Pi_{+}}
\nc{\pim}{\Pi_{-}}
\nc{\pih}{\Pi_{\lh}}
\nc{\jz}{J_{z}}
\nc{\jzh}{\jz^{\lh}}
\nc{\jzp}{\jz^{+}}
\nc{\jzm}{\jz^{-}}
\nc{\del}{\partial}
\nc{\dz}{\del_{z}}
\nc{\dzb}{\del_{\bar{z}}}
\nc{\az}{A_{z}}
\nc{\azb}{A_{\bar{z}}}
\nc{\g}{g^{-1}}
\nc{\dw}{\Delta_{W}}
\nc{\Ad}{{\mbox{Ad}}}
\nc{\ks}{Ka\-za\-ma-\-Su\-zu\-ki}
\nc{\KS}{\ks}
\nc{\ksm}{\ks\ model}
\rnc{\AA}{{\Bbb A}}
\nc{\BB}{{\Bbb B}}
\nc{\CC}{{\Bbb C}}
\nc{\PP}{{\Bbb P}}
\nc{\cpm}{\CC\PP(m)}
\nc{\cpn}{\CC\PP(n)}
\nc{\cp}[1]{\CC\PP(#1)}
\nc{\gmn}{G(m,m+n)}
\nc{\gmnk}{\gmn_{k}}
\nc{\cO}{{\cal O}}
\nc{\bcO}{\bar{\cO}}
\nc{\bO}{\bar{O}}
\nc{\oQ}{\overline{Q}}
\nc{\ie}{{\it i.e.~}}
\nc{\eg}{{\it e.g.~}}
\begin{document}
\global\parskip=4pt
\makeatother\begin{titlepage}
\begin{flushright}
{IC/98/197}
{ROM2F-98-39}
\end{flushright}
\vspace*{0.5in}
\begin{center}
{\LARGE\sc  D-strings in unconventional} \\ 
\vskip .3in
{\LARGE\sc  type I vacuum configurations}\\
\vskip .3in
\makeatletter
\begin{tabular}{c}
{\bf M. Bianchi}, \footnotemark 
\\ 
Universit\`a di Roma ``Tor Vergata'', Italy, \\
\end{tabular}

\begin{tabular}{c}
{\bf E. Gava}, \footnotemark 
\\ 
INFN, ICTP and SISSA, Trieste, Italy, \\
\end{tabular}

\begin{tabular}{c}
{\bf J.F. Morales}, \footnotemark
\\ 
SISSA, Trieste, Italy,  \\
\end{tabular}

\begin{tabular}{c}
{\bf K.S. Narain}\footnotemark
\\ ICTP, P.O. Box 586, 34014 Trieste, Italy\\
\end{tabular}
\end{center}
\addtocounter{footnote}{-1}%
\footnotetext{e-mail: bianchi@roma2.infn.it}
\addtocounter{footnote}{1}%
\footnotetext{e-mail: gava@he.sissa.it}
\addtocounter{footnote}{1}%
\footnotetext{e-mail: morales@roma2.infn.it}
\addtocounter{footnote}{1}%
\footnotetext{e-mail: narain@ictp.trieste.it}

\vskip .50in
\begin{abstract}
\noindent 
We determine the spectrum of D-string bound states in various classes of
generalized type I vacuum configurations with sixteen and eight supercharges.
The precise matching of the BPS spectra confirms the duality between
unconventional type IIB  orientifolds with quantized NS-NS
antisymmetric tensor and heterotic CHL models in $D=8$. A similar analysis puts
the duality between type II (4,0) models and type I strings
{\it without open strings} on a firmer ground. The analysis can be 
extended to type II (2,0) asymmetric orbifolds and their type I duals that 
correspond to unconventional $K3$ compactifications.
Finally we discuss BPS-saturated threshold corrections 
to the correponding low-energy effective lagrangians. In particular we show
how the exact moduli dependence of some ${F}^4$ terms in the eight-dimensional 
type II (4,0) orbifold is reproduced by the infinite
sum of D-instanton contributions in the dual type I theory. 

\end{abstract}
\makeatother
\end{titlepage}
\begin{small}
\end{small}

\setcounter{footnote}{0}

\section{Introduction}

The standard  $SO(32)$ heterotic - type I duality \cite{witdyn,pw}
requires  the existence of D-string bound
states \cite{witdbs}
in toroidally compactified type I theory that are mapped to 
fundamental heterotic winding states. In \cite{gmnt} the masses and 
multiplicities
of these  bound states carrying arbitrary quantum numbers was obtained by
studying the effective $O(N)$ gauge theory on the type I 
D-string worldsheet in the
infrared limit  where it flows to an orbifold conformal field theory.
In this paper we will extend these results to more general classes of dual
pairs with sixteen supercharges \cite{mbtor} 
and to some dual pairs with eight supercharges. We will consider two types
of models. In the first the type I side is modded by some orbifold group. 
In the second the orientifold $\Omega$ projection itself is modified 
by some orbifold group. In the first case the D-string effective action on the
type I side is just the original $O(N)$ gauge theory modded by further orbifold
group action. On the other hand in the second case one gets the D-string 
effective action by starting from type IIB D-strings and modding the open 
strings ending on them by the modified $\Omega$ projection. The resulting
D-string effective action thus turns out to be a different $Z_2$ projection 
of the $(8,8)$ $U(N)$ gauge theory living on the type IIB D-strings.

More precisely, in section 3, we will discuss the simplest instance
of type I duals to heterotic CHL models \cite{chl}. The former have been 
identified in \cite{mbtor,wittor} with unconventional type IIB orientifolds 
with quantized NS-NS antisymmetric tensor, 
introduced long time ago \cite{bpstor} and termed {\it BPS unorientifolds} 
in \cite{mbtor}. The
observed rank reduction of the gauge group in these generalized toroidal
compactifications admits a  gauge theoretic interpretation in terms of
non-commuting Wilson lines that represent the obstruction to
defining a {\it vector structure} in the vacuum gauge bundle on the torus 
\cite{lmst,mbtor,wittor,sese}. 

In section 4 we study type II (4,0) models {\it without D-branes}
that are dual to type I models {\it without open strings} \cite{mbtor,wittor,
gep,dp}.
The former arise from projections that include $(-)^{F_L}$ and remove all
massless R-R states to which D-brane usually couple \cite{cjp}. The latter
correspond to generalized $\Omega$-projections of the closed oriented type IIB
string that give rise to vanishing massless tadpoles in the transverse channel
Klein-bottle amplitudes thus preventing the introduction of D-banes and their
open-string excitations \cite{mbtor,wittor,gep,dp}. 
The duality relies on supersymmetry considerations (sixteen supercharges), 
matching of the 
massless spectra ($d=10-D$ matter vector multiplets) and S-duality invariance 
of the type IIB theory in $D=10$ that maps the $(-)^{F_L}$ 
action to the $\Omega$ worldsheet parity operation
through $(-)^{F_L}=S\Omega S^{-1}$
\cite{sen,mbtor}. We show that in the simplest $D=9$ case the D-string
bound-state spectrum in the type I models {\it without open strings}
agrees with the BPS spectrum of fundamental strings 
in type II (4,0) models. The exact matching of the
BPS spectra represents a precision test of the duality between these two
classes of models. 
Eventually we extend our analysis to dual pairs of theories with eight
supercharges corresponding to generalized orbifold/orientifold 
compactifications on $K3$. We will discuss in some detail the matching 
between a type II (2,0) model {\it without 
D-branes} and its dual type I model {\it without open strings} in the
$T^4/Z_2$ orbifold limit of $K3$ \cite{sen}.

In preparation for the discussions in sections 3 and 4, in section 2
we recall the
definition and explicitly compute the elliptic genera for some relevant 
orbifold CFT's corresponding to symmetric product spaces \cite{dmvv}.

As an application of the results in sections 3 and 4, in section 5, we compute 
D-string instanton contributions to BPS-saturated $F^4$ 
eight dimensional couplings
\cite{bfkov} and compare with the corresponding perturbative 
results on the dual side.     

Section 6 contains our 
conclusions and some directions for further investigation.

\section{Elliptic Genera and Symmetric Spaces}

In this section we compute the elliptic genera for the class of 
conformal field theories relevant in the discussion
of type I D-string bound states spectra 
and ``BPS-saturated'' threshold corrections.        
The dual pairs we will be interested in are contructed via 
orbifold/orientifold compactifications of the type IIB theory. 
The self-duality of the original type IIB string exchanges
the NS-NS and R-R antisymmetric tensors, mapping fundamental
string excitations with $N$ units of winding to bound states
of $N$ D-strings. The information about charges, masses and degeneracies
of these bound states is encoded in the elliptic genus of the low
energy gauge theory describing $N$ nearby D-strings in the generalized
type I vacuum. Following \cite{gmnt}, we compute this index in the 
infrared limit where the relevant gauge theories flow to $(8,0)$
orbifold conformal field theory in a symmetric target
space ${\cal M}^N/{\cal S}_N$.   
The details of the correspondence in each particular case will be
discussed in the next sections where the complete agreement between
the fundamental and bound-state spectra is established 
for theories with sixteen
supercharges, such as the type I duals of the CHL models in $D=8$ or the
type I duals of the type II (4,0) models in $D=9$ \cite{mbtor}, 
as well as for a class of models with eight supercharges \cite{sen}.

Given a two-dimensional CFT, the relevant elliptic genus for our
present discussion, is the 
partition function with right moving fermions in the odd spin structure: 
\bea
\chi(q,\bar{q}) = Tr_{R} (-)^{F_R} q^{L_0-c/24}\bar{q}^{\bar{L}_0-c/24}=
Tr_{R}(-)^{F_R}e^{-2 \pi\tau_2 H+2 \pi i R\tau_1 P_\sigma},    
\label{elliptic}
\eea
where $q=e^{2\pi i\tau}$ and $\tau$ is the genus-one 
worldsheet modulus. The constant $R$ is introduced for later convenience, 
$H$ and $P_\sigma$ are the Hamiltonian ($t$-evolution) and momentum along
the $\sigma$ direction respectively. The only dependence on $q$ of
$\chi(q,\bar{q})$ arises from the integration over the bosonic zero modes since
the trace in the odd spin structure receives contribution only from the right
moving ground states.  
The elliptic genus (\ref{elliptic}) thus effectively counts the number
of BPS states (ground states of the right moving supersymmetric sector),
or more precisely the difference between the number of fermionic and bosonic 
supersymmetric ground-states.

In the following, we will be interested in computing 
these elliptic genera for two dimensional CFT's 
obtained by modding out $N$ copies of a given 
world-sheet theory (usually itself an orbifold theory with orbifold group
$Z_K$) by the 
permutation group ${\cal S}_N$.  
We will be interested in the situations where the orbifold group $Z_K$ 
preserves some number, let us say ${\cal N}$,
of supersymmetries \footnote{Throughout the paper, we 
adhere to the standard, though somewhat confusing, notation of counting 
supersymmetries in four-dimensional units, so that four {\it real} supercharges
correspond to each supersymmetry.}.
The worldvolume theories for the D-strings under consideration 
are more like Green-Schwarz $\sigma$-models, where target-space supersymmetry 
is realized through the $Z_K$-invariant zero-modes of some periodic fermion 
which we will generically denote them by $S$ in the following.
The full orbifold group acting on $N$ copies of the world sheet theory is
therefore the semidirect product $G \equiv {\cal S}_N\ltimes Z_K^N$.

Let us briefly review how the orbifold elliptic genus  
is computed \cite{dhvw, dmvv, gmnt}. The Hilbert space for a 
non-Abelian orbifold
conformal field theory is built from the different twisted sectors 
associated to the conjugacy classes [{\bf g}]$=[(g,\alpha)]$ of
the orbifold group $G={\cal S}_N\ltimes Z_K^N$, 
(with $g \in {\cal S}_N$ and $\alpha\in Z_K^N$). In each sector we
project by the centralizer ${\cal C}_{\bf g}$ of the twist element ${\bf g}$ 
in $G$.     
The elliptic genus for this CFT is zero, 
{\it viz.}
$(2^{{\cal{N}}-1}-2^{{\cal{N}}-1})=0$, due to the
trace over the fermionic zero-modes associated
to the center of mass combination $S_{0}^1+S_0^2+\cdots S_0^N$.
That this combination is invariant under the orbifold group
is clear, since $S_{0}^i$ are
invariant under $Z_K$ and get permuted under ${\cal S}_N$. Indeed, these zero
modes generate the $2^{{\cal{N}}-1}$ bosonic and $2^{{\cal{N}}-1}$
fermionic components of short
BPS supermultiplets. Being interested in computing the degeneracies 
of such supermultiplets, only sectors with no additional fermionic zero-modes
will be relevant. 

Let us start by identifying these sectors. In order to achieve this goal it is
sufficient to consider the action of ${\cal S}_{N}$, 
since as already stated $Z_{K}^{N}$ 
acts trivially on the $S$ fields. 
A general conjugacy class $[g]$ in ${\cal S}_N$ is characterized
by partitions
${N_n}$ of $N$ satisfying $\sum n N_n=N$, where $N_n$ denotes the multiplicity
of the cyclic permutation $(n)$ in the decomposition
\be
[g]=(1)^{N_1}(2)^{N_2}\cdots (s)^{N_s}.
\ee   
The centralizer of an element in this conjugacy class takes the form
\be
{\cal C}_g = \prod_{n=1}^s {\cal S}_{N_n}\ltimes Z_n^{N_n}   
\quad ,
\ee
One immediately finds that if $[g]$ involves cycles of different lengths, say
$(n)^a$ and $(m)^b$ with $n \neq m$, the corresponding twisted sector
does not contribute to the elliptic genus. In order to see this, note that in 
this case there
are at least two sets of zero modes of $S$, which can be expressed,
by a suitable ordering of indices, as $(S^1+ S^2 +\cdots S^{na})$ and
$(S^{na+1}+\cdots S^{na+mb})$, where the two factors $(n)^a$ and $(m)^b$
act on the two sets of indices in the obvious way. These zero modes
survive the group projection because the centralizer of $g$ does not
contain any element that mixes these two sets of indices with each other,
thereby giving zero contribution to the elliptic genus. Thus we need only
to consider those sectors with $[g]=(L)^{M} $ where $N=LM $. 

In the $[g]=(L)^{M}$ case the centralizer is 
${\cal C}_g= {\cal S}_M \ltimes Z_L^M$.
>From the boundary condition along $\sigma$ it is clear that there are $M$
combinations of $S$'s that are periodic in $\sigma$. By suitable ordering,
they can be expressed as 
$\widehat{S}^{(\ell)} = \sum_{i=L\ell +1}^{L(\ell+1)}S^i$ for $\ell=0,\dots,
M-1$. These zero modes have to be projected by the elements $h$ in the
centralizer ${\cal C}_{\bf g}$. In particular, when $h$ is the generator of $Z_M
\subset {\cal S}_M \subset {\cal C}_{\bf g}$, 
it acts on the zero modes $\widehat{S}^{(\ell)} $ by a cyclic
permutation. It is clear, therefore, that only the center of mass
combination $\sum_{\ell=0}^{M-1} \widehat{S}^{(\ell)}$ 
is periodic along the $\tau$ direction.
Hence, this sector contributes to the elliptic genus. More generally any
$h=(h,\beta)\in {\cal C}_{\bf g} = {\cal S}_M \ltimes Z_L^M$ 
will satisfy the above criteria
provided $h= (M)\in {\cal S}_M$ and $\beta$ is some element in $Z_L^M$. 
The number of such elements $h$ is $(M-1)! \times L^M$.

The full orbifold group $G$ is specified by an element of ${\cal S}_N$ 
(discussed above)  together with an element of $Z_K^N$. Let
us consider a generic element $(g,\alpha)$. We denote by 
$e^{2\pi i t_\phi/K}$ the eigenvalue of a given field $\phi$
under the $Z_K$ action. The groups ${\cal S}_N$ and
$Z_K^N$ form a semi-direct product, since
${\cal S}_N$ acts as an automorphism in $Z_K^N$ by
permuting the various $Z_K$ factors. Denoting this action by
$g(\alpha)$, the semi-direct product is defined in the usual way: 
$(g,\alpha)\cdot (g',\alpha') = (gg', \alpha g(\alpha'))$. Twisted
sectors will be labeled by conjugacy classes in $G$. The relevant
sectors, for the elliptic genus computation, as discussed above, are the
conjugacy classes $[g]$ in ${\cal S}_N$ of the form $[g] = (L)^M$ with $N=LM$.
One can easily verify that the various classes in $G$ are labeled by
$([g], \alpha)$ with $\alpha =(\alpha_1,\alpha_2,\dots,\alpha_M)$
where each $\alpha_i$ is a representative in $Z_K^L$. It is easy to see
that these representatives $\alpha_i$'s are labelled by the diagonal
subgroup of $Z_K^L$.
Combining this with the condition we have found for $h$ we
may conclude that all the representatives $\alpha_i$'s must be equal (i.e.  all
$\alpha_i$'s must have the same eigenvalue under the diagonal subgroup of 
$Z_K^L$) 
in order for such an $h$ to exist in the centralizer 
of $([g],\alpha)$ in $G$. Different sectors are then characterized 
by a representative twist-element in the $\sigma$-direction
\bea
{\bf g}=(g;\alpha)=((L)^M;\alpha_1\cdot\alpha_2\cdots\alpha_M)
\label{g}
\eea
with $\alpha_1=\alpha_2 =\dots = 
\alpha_M=(e^{2\pi i \frac{t_\phi t_g}{K}},1,1,\dots 1)$ and
$t_g=0,1,\cdots K-1$ labelling the element of the diagonal subgroup
of $Z_K^L$. 
A sector twisted by a group element (\ref{g}) should be projected by 
the centralizer
\be
{\bf h}=(h;\beta)=
(Z_M\ltimes Z_L^M;\beta_1\cdot\beta_2\cdots \beta_M)) 
\label{h}
\ee
where $\beta \in Z_K^N$ satisfies $\alpha h(\alpha) = \beta g(\beta)$.  
The number of independent $\beta$'s of this kind is $K^M$ and
therefore the order of the centralizer of $([g],\alpha)$ is $M! L^M K^M$. 
The number of
elements $h$ in (\ref{h}) that give rise to a non-vanishing
trace is $(M-1)! L^M K^M$,
and therefore these are the relevant elements for the computation of
the elliptic genus.  However, not all the $h$'s of this form give different
traces. Indeed, if $h$ and $h'$ are in the same conjugacy class 
in ${\cal C}_{\bf g}$,
they will give the same trace. We can choose again a representative 
element ${\bf h}$. The orbifold group representatives      
(\ref{g}) and (\ref{h}) can be diagonalized with eigenvalues being given by:
\bea
{\bf g} &=& e^{2 \pi i (\frac{t_\phi t_g}{K L}+\frac{l}{L})}\nonumber\\
{\bf h} &=& e^{2 \pi i (\frac{t_\phi t_h}{K}+
s\frac{t_\phi t_g}{K N}+s\frac{l}{N}+\frac{r}{M})}
\label{gh}
\eea
with $(l,r)$ ($l=0, \cdots L-1$, $r=0, \cdots M-1$) 
denoting the $N=M\cdot L$ copies of
a generic field $\phi$. Different orbifold sectors are denoted 
by $(s, t_g, t_h)$, where $s=0,\dots, L-1$ labels  
the $Z_L$ elements in ${\bf h}$ and 
$t_g, t_h=0,\dots, K-1$ label
the $Z_K$ elements in the $\sigma$ and $\tau$ directions
respectively. 
It is easy to verify that the number of
elements in the centralizer ${\widehat{\cal C}}_{\bf h}$ in ${\cal C}_{\bf g}
$, 
for a relevant ${\bf h}$, is
$KML=KN$. As a result, the number of elements in the conjugacy class of
such a ${\bf h}$ in ${\cal C}_{\bf g}$ is 
$|{\cal C}_{\bf g}|/|{\widehat{\cal C}}_{\bf h}| = (M-1)!L^{M-1} K^{M-1}$.
Each of these classes appear with a
prefactor, given by the number of elements in the class divided
by the order of ${\cal C}_{\bf g}$, that is equal to $1/(KN)$.

We are now ready to compute the elliptic genus of the symmetric product of $
N$ copies of a worldsheet theory. We start with
the elliptic genus of a single copy of the theory:
\be
\chi^{osc}\equiv \prod_{\phi}\chi{\alpha_{\phi} \brack \beta_{\phi}}
\ee
where
\be
\chi{\alpha_{\phi} \brack \beta_{\phi}}=\prod_{n=1}^{\infty}
(1+e^{2 \pi i \beta_{\phi}}q^{n-1/2-\alpha_{\phi}})^{\epsilon_{\phi}}
\ee
is the oscillator-mode contributions of a generic, say right-moving, 
$\epsilon_{\phi}=-1(1)$ bosonic (fermionic)
field $\phi$ with boundary conditions given
by $\alpha_{\phi},\beta_{\phi}$ along $\sigma$ and $\tau$ directions 
respectively. We will omit in the following 
the zero mode contributions which are included at the end 
of the computation. 

We can write the contribution to the elliptic genus of a 
given ${\bf g,h}$ sector with eigenvalues given by (\ref{gh}) as
\be
\chi^{s,t_g, t_h}_{L, M}{\alpha_{\phi} \brack  \beta_{\phi}}(q)=
\prod_{r=0}^{M-1} \prod_{l=0}^{L-1} \prod_{n=0}^{\infty}
(1+e^{2 \pi i\beta_{\phi}(l,r)}
q^{n-1/2-\alpha_{\phi}(l)})^{\epsilon_{\phi}}
\ee
with
\bea
\alpha_\phi(l)&=& \alpha_\phi+\frac{t_\phi t_g}{K L}+\frac{l}{L}\nonumber\\
\beta_\phi(l,r)&=& \beta_\phi+\frac{r}{M}+\frac{t_\phi t_h}{K M}+
\frac{s}{N}(\frac{t_\phi t_g}{K}+l)\nonumber.
\eea
Performing the products over $r$ and $l$ yields
\be
\chi^{s,t_g, t_h}_{L, M}{\alpha_{\phi} \brack \beta_{\phi}}(q)=
\prod_{m=1}^{\infty}
(1+e^{2 \pi i \beta}
(q^{\frac{M}{L}} e^{-2 \pi i 
\frac{s}{L}})^{n-1/2-\alpha})^{\epsilon_{\phi}}
\label{chiab}
\ee 
in terms of the modified boundary conditions 
\bea
\alpha &=&
L(\alpha_\phi+\frac{1}{2})+t_\phi \frac{t_g}{K}+\frac{1}{2}\nonumber\\
\beta &=& 
M(\beta_\phi+\frac{1}{2})-s(\alpha_\phi+\frac{1}{2}) 
+t_\phi \frac{t_h}{K}+\frac{1}{2}  .
\label{modbc}
\eea 
Finally the zero-mode contributions to the elliptic genus depend on the
bosonic or fermionic nature of the field under consideration. For fermions
one finds:
\be
(2^{{\cal{N}}-1}-2^{{\cal{N}}-1}) 
\sqrt{\prod_{j=1}^{M-1}(1-e^{2 \pi i j/M})^{2{\cal{N}}}}
= (2^{{\cal{N}}-1}-2^{{\cal{N}}-1}) M^{\cal{N}} .
\label{ellfermi}
\ee
For bosonic fields there is a further distinction depending on the
compactness of the bosonic coordinate.
For $d$ compact bosons one has (in units of $\alpha^\prime =1/2$)
\be 
\sum_{(p,\bar{p})\in
\Gamma_{d,d}} (q^{\frac{M}{L}} e^{-2 \pi i \frac{s}{L}})^{p^2/2}
(\bar{q}^{\frac{M}{L}} e^{2 \pi i \frac{s}{L}})^{\bar{p}^2/2}  .
\label{ellcombos}
\ee
where ${\Gamma_{d,d}}$ is the even self-dual Lorentzian lattice of generalized
momenta, combining K-K momenta and winding modes.  
For $D$ non-compact bosons one gets 
\be 
L^{D} \tau_2^{-D/2} = L^D \int d^{^D}p \, (q\bar{q})^{p^2/2}
\label{ellnoncom}
\ee
We can now be more precise for some explicit choices of the 
initial worldsheet content. 

The first example we will consider is defined by the $Z_2$-orbifold 
of the worldvolume theory for $N$ copies of the heterotic 
Green-Schwarz string ($\phi=X^I, S^a,\chi^A$ with 
$\alpha_\phi=\beta_\phi=\frac{1}{2}$), 
where $Z_2$ is simply the GSO projection $\chi^A\rightarrow -\chi^A$. 
Specializing (\ref{chiab}-\ref{ellnoncom}) to the field content under 
consideration yields
\bea
\chi^I_N(q)&=&\frac{8-8}{\tau_2^4}\sum_{L,M}M^{-4}\frac{1}{2N}
\sum_{s=0}^{L-1}\sum_{t_g, t_h=0,1} 
\frac{\vartheta{\frac{t_g}{2} \brack  \frac{t_h}{2}}
(q^{\frac{M}{L}}e^{-2\pi i\frac{s}{L}})^{16}}
{\eta^{24}(q^{\frac{M}{L}}e^{-2\pi i \frac{s}{L}})}
\label{genusI}
\eea
where an overall factor $\frac{1}{N^8}$ has been included so that the longest 
string sector appears with unit normalization in accordance with the fact that
on $({\bf R}^8)^N$ there is only one fixed plane under the $Z_N$ action.

The second example start from a type IIB-like worldsheet theory 
($\phi=X^I, S^a, S^{\dot{a}}$) with $X^I, S^a$ periodic 
($\alpha_\phi = \beta_\phi=\frac{1}{2}$) and $S^{\dot{a}}$
antiperiodic ($\alpha_\phi=0,\beta_\phi=\frac{1}{2}$) fields.
In this case, (\ref{chiab}-\ref{ellnoncom}) yield
\bea
\chi^{\tilde{I}}_N(q)=\frac{8-8}{\tau_2^4}\sum_{L,M}M^{-4}\frac{1}{N}
\sum_{s=0}^{L-1}
\frac{\vartheta{\frac{L}{2}+\frac{1}{2}\brack  \frac{s}{2}+\frac{1}{2}}
(q^{\frac{M}{L}}e^{2\pi i \frac{s}{L}})^4}
{\eta^{12}(q^{\frac{M}{L}}e^{2\pi i \frac{s}{L}})}
\label{genusIIB}
\eea
where the same overall normalization $\frac{1}{N^8}$ has been included.

\section{Type I duals of heterotic CHL models}

In \cite{gmnt} the spectrum of D-string bound states for toroidal
compactifications of the type I theory was studied. Masses, multiplicities
and charges of bound states were read from the elliptic
genus of the effective $O(N)$ gauge theory describing $N$ 
nearby type I D-strings \cite{hetds,gmnt}:
\bea
S &=& \Tr \int d^{2}\sigma \; \left\{ -\frac{1}{4g^{2}}F^{2} + (DX_{I} )^{2} 
+ \Lambda_{\dot{a}} \slash{D}_R \Lambda^{\dot{a}} 
+ S_a \slash{D}_L S^a + \chi_{A} \slash{D}_R \chi^{A} 
\right. \nonumber \\
 &+& 
\left. g^2([X_{I},X_{J}])^{2} + 
g\Lambda^{\dot{a}}  \Gamma^I_{\dot{a}a}  [X_I, S^a]  
+  W^I_{\ell}\bar{\partial} X_I\chi^{A} (T^{\ell})_{AB}\chi^B \right\} \quad .
\label{Iaction}
\eea   
The fields transform in diverse representations of the gauge group
$O(N)$. $X$ and $S$ transform as second rank symmetric, traceless tensors,
while $\Lambda$ and $\chi$ transform in the adjoint and fundamental
representations respectively. The singlet (trace) parts of $X$ and $S$
represent the collective super-coordinates of the center-of-mass motion and
decouple from the rest. There is an $SO(8)_{\cal R}$, R-symmetry group, under
which $X$, $S$, $\Lambda$ and $\chi$ transform as an ${\bf 8}_V$ (labelled
by $I$), an ${\bf 8}_S$ (labelled by $a$), an ${\bf 8}_C$ (labelled by 
$\dot{a}$) and a singlet,
respectively. The $\chi$'s correspond to the Ramond ground states of open
strings stretching between D1- and D9-branes  \cite{pw} 
and transform under the vector
representations of both $SO(32)$ and $O(N)$. As reminded by the subscript of
the Dirac operators, $\Lambda$ and $\chi$ are negative chirality (right-moving)
worldsheet fermions while $S$ are positive chirality (left-moving) fermions.
Finally, $W^I_{\ell}$ are $SO(32)$ Wilson lines on the $I^{th}$ transverse
direction, $T^{\ell}$
being the $SO(32)$ generators in the vector representation. 
The presence of this term in the worldvolume
effective action can be deduced by explicitly computing the three-point 
function on the disk involving the vertex operators
for two $\chi$ fields and one bosonic coordinate $X$ in the
presence of $SO(32)$ Wilson lines $W^I_{\ell}$. In the canonical picture for the
external fields ($-{1\over 2}$ for the spinors and $-1$ for the bosons):
\be
\langle e^{-\frac{\varphi}{2}}S^+\sigma(0)\,e^{-\frac{\varphi}{2}}S^+
\bar{\sigma}(1)\,e^{-\varphi}\psi^I(\infty) 
\int dx \left( \del X^J + i p\cdot\psi\psi^J \right) (x) \rangle
\label{discamp}
\ee     
$\sigma, \bar{\sigma}$ represent the twist fields for the ND directions
of the transverse $X$ coordinates, $S^+$ the two dimensional longitudinal 
spin field and $\varphi$ the scalar arising from the superghost bosonization.
The $SO(32)\times O(N)$ group structure in (\ref{Iaction}) enters 
through the standard Chan-Paton (C-P) factor. From (\ref{discamp}) one easily
recognizes the last world-sheet coupling in (\ref{Iaction}).
In \cite{gmnt}, Wilson lines turned on the longitudinal
$\sigma$-direction of the D-string worlsheet were considered. In
that case the Wilson-line coupling reduces to a quadratic term
in the $\chi$ fields.  

As conjectured in \cite{hetds} and supported by the results of 
\cite{gmnt} the $O(N)$ gauge theory described by (\ref{Iaction})
flows in the infrared to an (8,0) orbifold conformal field theory 
given by 
the Green-Schwarz $\sigma$-model for $N$ copies of the heterotic string with
target space $(R^8)^N/{\cal S}_N$. It was shown how the longest-string sector 
of this orbifold theory reproduces the charges, masses 
and degeneracies required by the duality relation with the fundamental
heterotic BPS states.
In this section we show how a similar analysis can be extended to more general 
vacuum configurations with sixteen supercharges. 
In particular we consider the type I dual \cite{mbtor,wittor} of 
a CHL model \cite{chl} in $D=8$, but it will be clear from the discussion
that the arguments are rather general and apply to many, if not all,
unconventional heterotic - type I dual pairs. 

Let us consider, for example, a dual pair with gauge group $SO(16)$ 
in $D=8$. From the type I perspective, models of this kind were constructed 
long time ago 
\cite{bpstor} as open-string descendants of the type IIB string on tori with
non-vanishing but quantized NS-NS antisymmetric tensor $B^{^{NS}}$. 
The presence of a 
quantized $B^{^{NS}}$ in these type I models, more concisely
termed {\it BPS unorientifolds} in \cite{mbtor}, 
has been identified with the obstruction to defining a vector structure 
in the vacuum gauge bundle or
equivalently with the presence of non-commuting Wilson lines on the torus
\cite{mbtor,wittor}.
In $D=8$, the dual heterotic CHL model 
\cite{chl} can be constructed by turning on 
non-commuting Wilson lines on the two-torus of a conventional
$SO(32)$ heterotic string compactification \cite{lmst}. Non-commuting Wilson
lines  effectively modify the boundary conditions 
of the $\chi$ fields representing the $SO(32)$ algebra. 
We can therefore realize them as non-abelian orbifolds $T^2/G$.
Let us denote the generators of $G$ as $g_1, g_2$. In order
to keep an $SO(16)$ component of the gauge group we 
decompose the 32 heterotic fermions $\chi^A$ into 
two groups\footnote{For lack of a better symbol, we 
continue to label the fermions in the two subsets with $A=1,\dots 16$.
We hope this would not cause confusion with the initial range of the same 
label $A=1,\dots 32$.} of 16 each
$\chi_{(1)}^A$ and $\chi_{(2)}^A$
 and choose
\bea
g_1 &:& X_8\rightarrow X_8+\pi R_8; 
\quad \chi_{(1)}^A \rightarrow +\chi_{(1)}^A 
\quad \chi_{(2)}^A \rightarrow -\chi_{(2)}^A  \nonumber\\
g_2 &:& X_9\rightarrow X_9+\pi R_9
\quad \chi_{(1)}^A \rightarrow +\chi_{(2)}^A 
\quad \chi_{(2)}^A \rightarrow +\chi_{(1)}^A \quad .
\label{eta}
\eea
The first action in (\ref{eta}) breaks $SO(32)$ to $SO(16)^2$ at level $k=1$.
The second projects on the diagonal $SO(16)$ gauge group at level $k=2$. 
The half shifts ensure that no new massless gauge bosons arise in the twisted
sectors. Up to the original GSO projection on the world-sheet
fermions, twisted sectors corresponds to the conjugacy classes 
$[1],[g_1],[g_2],[g_1 g_2]$ of $G$. 
In each of these sectors we project by the centralizer ${\cal C}_g$, 
that only in the untwisted sector ($[1]$)
coincides with the full $G$, because $g_1$ and $g_2$ do not 
commute.

Following \cite{mbtor} we can write the full CHL partition function 
as a sum of the contributions from the four sectors 
$[1],[g_1],[g_2],[g_1 g_2]$ that reads
\bea
Z_{(1,1)}&=&\frac{1}{8 \tau_2^4}\frac{\bar{\cal Q}}{\bar{\eta}^8}(\bar{q})
\frac{\vartheta_3^{16}+\vartheta_4^{16}+\vartheta_2^{16}}{\eta^{24}}(q)
\sum_{P\in \Gamma_{2,2}}
q^{{p^2}/{2}}\bar{q}^{{\bar{p}^2}/{2}}
\label{11}\\
Z_{(1,g_1)}&=&\frac{1}{4 \tau_2^4}\frac{\bar{\cal Q}}{\bar{\eta}^8}(\bar{q})
\frac{\vartheta_3^{8}\vartheta_4^{8}}{\eta^{24}}(q)
\sum_{P\in \Gamma_{2,2}}e^{i \pi P\cdot V_8}
q^{{p^2}/{2}}\bar{q}^{{\bar{p}^2}/{2}}
\label{1g1}\\
Z_{(1,g_2)}&=&\frac{1}{4 \tau_2^4}\frac{\bar{\cal Q}}{\bar{\eta}^8}(\bar{q})
\frac{\vartheta_3^{8}\vartheta_4^{8}}{\eta^{24}}(q)
\sum_{P\in \Gamma_{2,2}}e^{i \pi P\cdot V_9}
q^{{p^2}/{2}}\bar{q}^{{\bar{p}^2}/{2}}
\label{1g2}\\
Z_{(1,g_1g_2)}&=&\frac{1}{4 \tau_2^4}\frac{\bar{\cal Q}}{\bar{\eta}^8}(\bar{q})
\frac{\vartheta_3^{8}(q^2)+\vartheta_2^{8}(q^2)}{\eta^{8}(q)\eta^8(q^2)}
\sum_{P\in \Gamma_{2,2}} e^{i \pi P\cdot V_{89}} 
q^{{p^2}/{2}}\bar{q}^{{\bar{p}^2}/{2}}
\label{1g1g2}\\
Z_{(g_1,1)}&=&\frac{1}{4 \tau_2^4}\frac{\bar{\cal Q}}{\bar{\eta}^8}(\bar{q})
\frac{\vartheta_3^{8}\vartheta_2^{8}}{\eta^{24}}(q)
\sum_{P\in \Gamma_{2,2}+V_8}
q^{{p^2}/{2}}\bar{q}^{{\bar{p}^2}/{2}}
\label{g11}\\
Z_{(g_1,g_1)}&=&\frac{1}{4 \tau_2^4}
\frac{\bar{\cal Q}}{\bar{\eta}^8}(\bar{q})
\frac{\vartheta_4^{8}\vartheta_2^{8}}{\eta^{24}}(q)
\sum_{P\in \Gamma_{2,2}+V_8}e^{i P\cdot V_8}
q^{{p^2}/{2}}\bar{q}^{{\bar{p}^2}/{2}}
\label{g1g1}\\
Z_{(g_2,1)}&=&\frac{1}{4 \tau_2^4}\frac{\bar{\cal Q}}{\bar{\eta}^8}(\bar{q})
\frac{\vartheta_3^{8}\vartheta_2^{8}}{\eta^{24}}(q)
\sum_{P\in \Gamma_{2,2}+V_9} q^{{p^2}/{2}}\bar{q}^{{\bar{p}^2}/{2}}
\label{g21}\\
Z_{(g_2,g_2)}&=&\frac{1}{4 \tau_2^4}
\frac{\bar{\cal Q}}{\bar{\eta}^8}(\bar{q})
\frac{\vartheta_4^{8}\vartheta_2^{8}}{\eta^{24}}(q)
\sum_{P\in \Gamma_{2,2}+V_9}e^{i \pi P\cdot V_9}
q^{{p^2}/{2}}\bar{q}^{{\bar{p}^2}/{2}}
\label{g2g2}\\
Z_{(g_1g_2,1)}&=&\frac{1}{4 \tau_2^4}\frac{\bar{\cal Q}}{\bar{\eta}^8}(\bar{q})
\frac{\vartheta_3^{8}(\sqrt{q})+\vartheta_4^{8}(\sqrt{q})}
{\eta^{8}(q)\eta^8(\sqrt{q})}
\sum_{P\in \Gamma_{2,2}+V_{89}}
q^{{(p^2}/{2}}\bar{q}^{{\bar{p}^2}/{2}}
\label{g1g21}\\
Z_{(g_1g_2,g_1g_2)}&=&\frac{e^{i\pi/3}}{4\tau_2^4}
\frac{\bar{\cal Q}}{\bar{\eta}^8}(\bar{q})
\frac{\vartheta_3^{8}(-\sqrt{q})+\vartheta_4^{8}(-\sqrt{q})}
{\eta^{8}(q)\eta^8(-\sqrt{q})}
\sum_{P\in \Gamma_{2,2}+V_{89}}e^{i \pi P\cdot V_{89}}
q^{{p^2}/{2}}\bar{q}^{{\bar{p}^2}/{2}}
\label{g1g2g1g2}
\eea
where ${\cal Q}=(\vartheta_3^{4}-\vartheta_4^{4}-\vartheta_2^{4})/\eta^4$
is the ubiquituous ``supersymmetric character" that vanishes in virtue of       
Jacobi's {\it aequatio identica satis abstrusa} and ${\Gamma_{2,2}}$ is the 
even self-dual Lorentzian lattice of generalized
momenta. As usual, the first subscript in the above amplitudes
refers to the twist in the $\sigma$ direction and the
second to the twist in the $\tau$ direction. We choose the order two 
``geometric" shifts generated by $V_8,V_9,V_{89}=V_8+V_9$ with 
$V_i=(v_i,\bar{v}_i)$ such that $v_8=-\bar{v}_8=R_8/2$ and 
$v_9=-\bar{v}_9=R_9/2$. The $\frac{1}{2}$ BPS spectrum is defined 
by replacing ${{\cal Q}}/{\eta^8}$ with
its ground states $8_B-8_F$ and imposing the level matching
condition
\bea
\frac{1}{2}p^2-\frac{1}{2}\bar{p}^2=m_8 n_8+ m_9 n_9=N_L-c_L
\label{chllm}
\eea
where $m_i$ and $n_i$ (integers or half integers depending on the orbifold 
sector) represent the winding and momentum modes of the fundamental
strings and $c_L$ the zero point energies in the corresponding sector.
In particular for a string wrapping $N$ times around a single direction 
the level matching condition (\ref{chllm}) simply reduces to 
$nN=(N_L-c_L)$ and the partition function for such states
can be read off from the $N=1$ partition function after replacing $q$ with 
$q^{\frac{1}{N}}$.    

We can now compare the above BPS spectrum with the BPS spectrum of the
conjectured dual type I model \cite{mbtor,wittor}.
The spectrum of $\frac{1}{2}$-BPS states in this unconventional type I 
toroidal compactification consists in the Kaluza-Klein (K-K) excitations of the 
massless type I string states and the
D-string windings around the eigth and nine directions. 
The former are encoded in the one-loop partition function
that involves the Klein-bottle ${\cal K}$, annulus ${\cal A}$ and 
M\"obius-strip ${\cal M}$
amplitudes in addition to half the {\it parent} type IIB torus amplitude.
For the $SO(16)$ type I model in $D=8$ \cite{bpstor,mbtor} under consideration 
the various amplitudes read
\bea
{\cal T} &=& \frac{1}{2} \frac{{\cal V}_8}{4 \pi^2 \alpha^\prime}
\int_{\cal F}
\frac{d^2\tau}{\tau_2^2}\frac{1}{\tau_2^3}\frac{|{\cal Q}|^2}{|\eta|^{16}}
(\tau,\bar{\tau})
\sum_{p\in\Gamma_{2,2}} 
q^{{p^2}/{2}}\bar{q}^{{\bar{p}^2}/{2}}\label{torus}\\
{\cal K} &=& \frac{1}{2} \frac{{\cal V}_8}{4 \pi^2 \alpha^\prime}
\int_0^\infty \frac{dt}{t}\frac{1}{t^4}\frac{{\cal Q}}{\eta^{8}}(2it)
\sum_{p\in\Gamma_{_{KK}}} 
e^{-\pi t p^2}\label{klein}\\
{\cal A} &=& \frac{N^2}{2} \frac{{\cal V}_8}
{4 \pi^2 \alpha^\prime}\int_0^\infty
\frac{dt}{t}\frac{1}{t^4}\frac{{\cal Q}}{\eta^{8}}(\frac{it}{2})
\sum_{\epsilon}\sum_{p\in\Gamma_{_{KK}}+\epsilon} 
e^{-\pi t p^2}\label{annulus}\\
{\cal M} &=& \frac{N}{2} \frac{{\cal V}_8}
{4 \pi^2 \alpha^\prime}\int_0^\infty
\frac{dt}{t}\frac{1}{t^4}\frac{{\cal Q}}{\eta^{8}}(\frac{it}{2}+\frac{1}{2})
\sum_{\epsilon}\sum_{p\in\Gamma_{_{KK}}+\epsilon}\gamma_\epsilon 
e^{-\pi t p^2}\label{moebius}
\eea
where ${\cal V}_8$ is a regularizing volume and
${\Gamma_{2,2}}$ is a left-right symmetric even self-dual Lorentzian 
lattice of generalized momenta, combining K-K momenta and winding modes.  
The condition of left-right symmetry results in a quantization condition 
for the NS-NS antisymmetric tensor $B^{^{NS}}$ \cite{bpstor}. 
The corresponding restriction on the closed-string states flowing
in the transverse channel halves the rank of the C-P group 
and induces the presence of order-two shifts $\epsilon$  
in the two dimensional lattice of K-K momenta $\Gamma_{KK}$. 
The $Z_2$ phases, \ie signs, $\gamma_\epsilon$ and the C-P multiplicities $N$
are constrained by R-R charge neutrality. Indeed,
tadpole cancellation implies $\sum_\epsilon \gamma_\epsilon=-2$ and 
$N=16$. The four possible choices for
$\gamma_\epsilon$ correspond to three inequivalent $SO(16)$ and one $Sp(16)$
type I models \cite{wittor}. The spectrum of perturbative BPS states
can be read off 
after putting the supersymmetric part in one of its ground states. 
Up to the resulting overall multiplicity $8_B-8_F$, one finds
8 $SO(16)$ singlets from the K-K reduction 
of the supergravity sector
(\ref{torus},\ref{klein}), three {\bf 120} antisymmetric tensors 
from the sectors
with K-K momenta shifted by $\epsilon=(0,0),(0,\frac{1}{2}),
(\frac{1}{2},0)$
with $\gamma_{\epsilon}=-1$ and finally one symmetric tensor  
({\bf 135}+{\bf 1}) from the sector with K-K momenta
shifted by $\epsilon=(\frac{1}{2}, \frac{1}{2})$ with $\gamma_{\epsilon}=+1$.

We can now compare the type I perturbative spectrum with its
heterotic CHL counterpart defined by the fundamental strings with no-winding
modes $m_8=m_9=0$. Recalling that the $V$'s in (\ref{g11}-\ref{g1g2g1g2})
represent shifts in the winding heterotic modes, states with $m_8=m_9=0$
only arise in the untwisted sector 
(\ref{11}-\ref{1g1g2}). Imposing the level matching condition
(\ref{chllm}) ($q^0$ order in the expansions of (\ref{11}-\ref{1g1g2})) 
gives rise to 8+{\bf 120} supermultiplets for $n_8,n_9$ both even,
{\bf 120} supermultiplets for one $n$ even and the other odd, and {\bf 136}
supermultiplets for $n_8,n_9$ both odd. The spectrum of 
heterotic BPS states with zero winding coincides with the previously
found type I perturbative spectrum after the identifications $R_8^{H}=2 R_8^I$
and $R_9^{H}=2 R_9^I$.
   
The winding modes of the fundamental strings, on the other hand,
are mapped to bound states of D-strings in the $SO(16)$ type I model 
under consideration. For example the basic $N=1$ unit (after the rescaling
of the internal radius) of D-string winding in the $8^{th}$ direction will 
be represented in the dual theory by the fundamental heterotic strings in the
sectors twisted by an element in $[g_1]$ or in $[g_1 g_2]$. 
In order to identify the relevant D-string gauge theory in the dual type I
model we should combine the $\Omega$ projection with the $g_1$, $g_2$
actions. For example for a single winding around the $8^{th}$ direction
the corresponding worldvolume theory is defined in terms of
the $\Omega g_1$-invariant fields  
\begin{eqnarray}
X^I,S^{\dot{a}},\chi^A_{(1)}&{\rm for}& P_\sigma~~ {\rm even}
\nonumber\\
\chi^A_{(2)}&{\rm for}& P_\sigma~~ {\rm  odd}
\label{contentg1}
\end{eqnarray}  
After defining a new longitudinal variable $\tilde{\sigma}=2 \sigma$, 
we are left with
a CFT in terms of the $\tilde{\sigma}$-periodic $X,S,\chi_{(1)}$ fields
and the $\tilde{\sigma}$-antiperiodic $\chi_{(2)}$ fields. 
Similarly for the $g_1 g_2$ $Z_4$-action we have a two dimensional
free theory in terms of
periodic $X,S$ fields and $\chi$ fields satisfying the 
boundary conditions 
$\chi_{\pm}^A(\tilde{\sigma}+1)=\pm i\,\chi_{\pm}^A(\tilde{\sigma})$.
For $N$ D-strings we should deal with the same $O(N)$ effective action
(\ref{Iaction}) as in the standard type I theory now with the $\chi$
fields satisfying these new boundary conditions. The arguments of
\cite{gmnt} directly apply to this D-string to show that the infrared
limit of the gauge theory is governed by an orbifold conformal theory,
defined by $N$ copies of the $N=1$ D-strings above moving on
the symmetric space $({\cal M})^N/{\cal S}_N$. The manifold ${\cal M}$ is
defined by the $Z_2$ orbifold $R^7\times (S^1)/g_2$ with $g_2$ defined
as before. The symmetrization of this orbifold should be understood as follows:
since the actions of $g_1$ and $g_2$ do not commute, a sensible $g_2$ action
can be defined only in a sector with $N$ even. More precisely the symmetric
space can be understood as the symmetrization of $N$ copies of the conformal
theory defined by the $N=1,2$ sectors. 

We can now apply formula (\ref{genusI}) and check that the longest string
sector ($M=1, L=N$) precisely yields 
the multiplicities of the D-string bound states wrapped
$N$ times around the $8^{th}$ direction. 
We start with the amplitudes that are $g_2$-twisted in neither two 
world-sheet directions. 
For the CFT defined by $N$ copies
of the $[g_1]$-twisted heterotic strings 
($\alpha_\phi=0, \beta_\phi=\frac{1}{2}$ for $\phi=\chi^A_{(1)}$,
$\alpha_\phi=\beta_\phi=\frac{1}{2}$ otherwise, $Z_K\equiv Z_2^{GSO}$) 
we find
\bea
\chi^I_N(q)&=&\frac{8-8}{\tau_2^4}\frac{1}{2N}
\sum_{s=0}^{N-1}\sum_{t_g, t_h=0,1} 
\frac{1}{\eta^{24}(\tilde{q})}
\vartheta{\frac{t_g}{2} \brack  \frac{t_h}{2}}^8
(\tilde{q})
\vartheta{\frac{t_g}{2}+\frac{N}{2} \brack \frac{t_h}{2}+\frac{s}{2}}^8
(\tilde{q})
\label{genusIg1}
\eea
with $\tilde{q}\equiv q^{\frac{1}{N}}e^{-2\pi i \frac{s}{N}}$.
Starting with $N$ copies of a heterotic string in the $[g_1 g_2]$
sector ($\alpha_\phi=\frac{1}{4}(\frac{3}{4}),
\beta_\phi=\frac{1}{2}$ for $\phi=\chi^A_{+}(\chi^A_-)$,
$\alpha_\phi=\beta_\phi=\frac{1}{2}$ otherwise) yields
\bea
\chi^I_N(q)=\frac{8-8}{\tau_2^4}\frac{1}{2N}
\sum_{s=0}^{N-1}\sum_{t_g, t_h=0,1} 
\frac{1}{\eta^{24}(\tilde{q})}
\vartheta{\frac{t_g}{2}+\frac{3N}{4} \brack  \frac{t_h}{2}+ \frac{3s}{4}}^8
(\tilde{q})
\vartheta{\frac{t_g}{2}+\frac{N}{4} \brack \frac{t_h}{2}+\frac{s}{4}}^8
(\tilde{q})
\label{genusIg1g2}
\eea

For different values of $N$, (\ref{genusIg1}) 
yields, in addition to the $(8_B-8_F)$ factor
associated to the center-of-mass fermionic zero-modes, 
degeneracies that are determined by the expansion of
\bea
{\rm s \quad odd, 
\quad N \quad odd}& &\frac{1}{\eta^{24}(\tilde{q})}
\vartheta_4^8(\tilde{q})\vartheta_2^8
(\tilde{q})
\sum_{P\in \Gamma_{1,1}}\tilde{q}^{p^2/2}\bar{\tilde{q}}^{\bar{p}^2/2}\\
{\rm s \quad even, 
\quad N \quad odd}& &\frac{1}{\eta^{24}(\tilde{q})}
\vartheta_3^8(\tilde{q})\vartheta_2^8
(\tilde{q})\sum_{m_9 \in Z} \tilde{q}^{p^2/2}\bar{\tilde{q}}^{\bar{p}^2/2}\\
{\rm s \quad odd, 
\quad N \quad even}& &\frac{1}{\eta^{24}(\tilde{q})}
\vartheta_4^8(\tilde{q})\vartheta_3^8
(\tilde{q})\sum_{m_9 \in Z} \tilde{q}^{p^2/2}\bar{\tilde{q}}^{\bar{p}^2/2}\\
{\rm s \quad even, 
\quad N \quad even}& &\frac{1}{\eta^{24}(\tilde{q})}
(\vartheta_2^{16}(\tilde{q})+\vartheta_2^{16}(\tilde{q})+\vartheta_2^{16}
(\tilde{q}))
\sum_{m_9 \in Z} \tilde{q}^{p^2/2}\bar{\tilde{q}}^{\bar{p}^2/2}\\
\eea
The sum over $s$ projects onto states which satisfy the heterotic
level matching condition (\ref{chllm}). 
Indeed for odd $N$ the above formula precisely matches with 
the degeneracies of the
$g_1$-twisted heterotic sectors (\ref{g11},\ref{g1g1}) while for
$N$ even it matches with the multiplicities coming from the untwisted sectors
(\ref{11}, \ref{1g1}). 

Similarly, starting with the partition function for
$N$ copies of the $[g_1g_2]$-twisted heterotic string given in
(\ref{genusIg1g2}) one can easily show that they yield 
the degeneracies coming from (\ref{g1g21}) and its T-modular
transform for $N$ odd and $s$ even and odd respectively; 
and (\ref{1g1g2},\ref{11}) for $N$ even and $s$ odd and even respectively. 

Finally for $N$ even we are allowed to perform the $g_2$-projection. This 
will reproduce the multiplicities in (\ref{1g2},\ref{g21},\ref{g2g2})
in an obvious way since, as we have just
discussed, $g_2$-untwisted sectors in both $\sigma$ and $\tau$ directions 
are given by the untwisted sector of the heterotic string. 

\section{Type IIB orbifold/orientifold dual pairs}

The other classes of models we are going to discuss 
in this section correspond to ${\cal N} = (4,0)$ and ${\cal N} = (2,0)$ 
supersymmetric\footnote{The even more
confusing notation ${\cal N}=({\cal N}_L,{\cal N}_R)$ 
corresponds to the splitting 
${\cal N} = {\cal N}_L+{\cal N}_R$, where ${\cal N}_L$, respectively 
${\cal N}_R$, counts, in four-dimensional 
units, the ``target-space" supersymmetries
associated to left-moving, respectively right-moving, spin fields.} 
vacuum configurations of type IIB string \cite{fk}. These type II models
are dual to type I models without open strings \cite{mbtor}.

\subsection{ Type IIB on $T^d/(-)^{F_L}\sigma_V$ vs. 
Type IIB on $T^d/\Omega\sigma_V$} 

We will first consider $(4,0)$ models,
which are obtained as asymmetric orbifolds \cite{dhvw} of $d$-dimensional
tori $T^d$, where
the orbifold group is generated by $(-)^{F_L}\times \sigma_V$.
$F_L$ is the spacetime left-moving fermion number while
$\sigma_V$ is a shift of order two in the $\Gamma_{d,d}$ lattice of 
generalized momenta
of the torus \cite{mbtor,fk}. As a result of the orbifold 
projection,
the left-moving supercharges are projected out in the untwisted 
sector. Due to the shift in the momentum lattice, no
supercharge appears in the twisted sector.
As remarked in the introduction, the orbifold projection removes all
R-R states. Thus type II $(4,0)$ models are effectively type II models 
without D-branes in that, unlike the standard ones \cite{cjp}, D-branes 
in type II
$(4,0)$ models do not give rise to BPS saturated states since there is no R-R 
counterpart of the NS-NS coupling to the graviton and dilaton.
This line of resoning leads one to conclude that, neglecting the states 
associated to wrapping the NS 5-brane that only play a role
in low enough dimensions ($D\leq 4$), the spectrum of BPS states
is completely perturbative much in the same way as in heterotic models.
 
The BPS perturbative spectrum for the type II $(4,0)$  models
can be read off from the elliptic genus 
\be
Tr_R (-)^{F_R}e^{-2\pi \tau_2 H + 2\pi i R\tau_1 P_\sigma} \, ,
\label{ellwdb}
\ee
where the trace is
computed in the right-moving Ramond sector: indeed in this case we are  
effectively setting the left-moving
oscillator number $N_L$ to its ground state value $N_L=0$. 
We are then left with the following
non-trivial contributions from the different orbifold sectors \cite{mbtor}:
\begin{eqnarray}
Z_{+-}&=&\frac{(8_v-8_s)}{\tau_2^{4-d/2}}
        \frac{\vartheta_2(q)^4}{\eta(q)^8}
        \sum_{P\in \Gamma_{d,d}}e^{2\pi i V\cdot P}
         q^{p^2/2}\bar{q}^{\bar{p}^2/2}\label{theta2}\\
Z_{-+}&=&\frac{(8_v-8_s)}{\tau_2^{4-d/2}}
        \frac{\vartheta_4(q)^4}{\eta(q)^8}
        \sum_{P\in \Gamma_{d,d}+V} 
        q^{p^2/2}\bar{q}^{\bar{p}^2/2}\label{theta4}\\
Z_{--}&=&-e^{\pi i V\bar{V}}\frac{(8_v-8_s)}
          {\tau_2^{4-d/2}}\frac{\vartheta_3(q)^4}{\eta(q)^8}
        \sum_{P\in \Gamma_{d,d}+V}e^{2\pi i V\cdot P}
         q^{p^2/2}\bar{q}^{\bar{p}^2/2} \quad .
\label{theta3}
\end{eqnarray}

In the above expressions the first subscript $\pm$ refers to the
twists in the $\sigma$ direction whereas the second to the twist in 
the $\tau$ direction. The factor $8_v-8_s$ comes from 
the right-moving fermionic zero-modes and counts the 8 bosons and 8
fermions of the short ${\cal N}=4$ supermultiplets, 
$V$ is the shift vector generating the action
of $\sigma_V$. Finally $P=(p,\bar{p})$ are the generalized momenta
in the compact directions.

The only gauge fields in these models
are those coming from the K-K reduction of
the ten dimensional metric $G_{MN}$ and NS-NS antisymmetric tensor
$B_{MN}^{NS}$, since, as discussed above,
there are no gauge fields arising from the R-R sector.
The corresponding charges belong to the (shifted) lattices appearing
in eqs.(\ref{theta2}) to (\ref{theta3}).  

For simplicity we will restrict the following discussion
to the 9-dimensional case of compactification on a circle
of radius $R$. In this case
the charges associated to the 9-dimensional gauge
fields $B_{\mu 9}^{NS}$ and $G_{\mu 9}$ are the winding
number $N$ and K-K momentum $k/R$, respectively. We will also take
the ``geometric" shift $\sigma_V$ to be genearted by the vector
$V=(v,\bar v)=(R/2,-R/2)$.  The level matching
condition then reduces to
\begin{equation}
kN=N_L-c_L ,
\label{LM}
\end{equation}
where $k$ is integer (half-integer) in the
untwisted (twisted) sector and $c_L$ denotes, as before,
the zero point energy in the left moving sector.
The multiplicity of these BPS states for given
$k$ and $N$ is defined by the coefficient of $q^{N_L}$
in the expansion of  (\ref{theta2}-\ref{theta3}), with $N_L$ given by
(\ref{LM}). 

The conjectured S-duality of the ten-dimensional
type IIB string
exchanges the NS-NS  and R-R antisymmetric tensors and 
maps winding modes of the fundamental string
into D-string winding modes. We then conclude that 
in the system of $N$ type I D-strings, each one wrapped once
around the circle, there should exist bound states  
carrying a given  K-K momentum $k$, with mass and degeneracy    
given by the above relations through the duality map.     
As discussed in \cite{gmnt}, in the context of type I - heterotic
duality, these data about the
bound states are encoded in the elliptic genus of the
effective gauge theory describing the dynamics of $N$
nearby type I D-strings.

In order to identify the effective gauge theory
governing the dynamics of type I D-strings 
in the present situation, notice that, as shown in \cite{sen}, 
S-duality maps $(-)^{F_L}$ into the world-sheet parity operator $\Omega$.
In $D=10$ the two quotient theories are vastly different. On the one hand,
projecting the type IIB theory by $(-)^{F_L}$ gives the type IIA theory since
the twisted sector of the orbifold provides the extra (opposite-chirality)
supercharges needed to restore (non-chiral) maximal supersymmetry. 
On the other hand,
projecting the type IIB theory by $\Omega$ gives the type I theory since
the twisted sector of the {\it parameter-space orbifold} 
\cite{ascar,hor,bs}, now commonly termed {\it orientifold},  
provides the $SO(32)$ C-P multiplicities
in terms of open-string excitations of the D9-branes needed to soak up the 
non-vanishing R-R charge of the O9-planes associated to the Klein-bottle 
$\Omega$-projection \cite{cjp}.

Nevertheless, accompanying the $Z_2$-projection by a shift in the
compactification torus results in dual pairs in lower dimensions
\cite{sen,mbtor,fk}. Closely following the analyses in \cite{gmnt}
and in section 2, the relevant gauge theory
will be obtained by projecting the $U(N)$ gauge theory on the world-sheet
of type IIB D-strings \cite{witdbs} 
onto $\Omega \sigma_V$ invariant fields. The K-K momentum of the
fundamental string corresponds to  $P_\sigma$ in the D-string system,
and therefore, $\sigma_V$, which is 
$+1$ or $-1$ depending on whether the K-K momentum is
even or odd, corresponds to (anti-)periodic boundary conditions
along the $\sigma$-direction on the D-string world-sheet.
Recalling that in the action of $\Omega$ 
on the $U(N)$ C-P factors there is
a relative sign between the transverse and longitudinal
degrees of freedom,  it is easy to obtain
the field  content resulting from the $\Omega\sigma_V$
projection. Let us denote by $A_\alpha$, $X^I$, $S^a$
and $S^{\dot{a}}$ the gauge fields,
transverse scalars, left- and right-moving 
fermions respectively, of the $U(N)$ type IIB D-string 
system. The indices $I,a,\dot{a}$ refer to the ${\bf 8}_v$,
${\bf 8}_s$ and ${\bf 8}_c$ representations of the $SO(8)$
R-symmetry group. The result of the $\Omega\sigma_V$ projection
is 
\begin{eqnarray}
X_I^{(+)}, S_{\dot{a}}^{(+)}, S_{a}^{(-)}, A_{\alpha}^{(-)} &{\rm for}& 
P_\sigma~~ {\rm even} \nonumber\\
X_I^{(-)}, S_{\dot{a}}^{(-)}, S_{a}^{(+)}, A_{\alpha}^{(+)} &{\rm for}& 
P_\sigma~~ {\rm  odd}
\label{content}
\end{eqnarray}
where $(+)$ and $(-)$ denote the symmetric and adjoint representations
of $O(N)$ respectively. 

Given these preliminar observations, one can follow the approach of
\cite{gmnt} where the elliptic genus of the gauge theory was computed by going
to the infrared limit. In this limit the theory has been shown to flow to a
superconformal orbifold field theory. Indeed, the charged fields
get massive and can be integrated out leaving a free field theory in
terms of the diagonal fields 
$X_I^t$, $S_a^t$ and $S_{\dot{a}}^t$, ($ t=1,\dots ,N$) in (\ref{content}).
Finally, modding out by the residual Weyl symmetry group, we are
left with an orbifold theory with target space
$({\bf R}^8)^N/{\cal S}_N$. 

Let us first analyze the free $N=1$ case. It is convenient to define  
a new variable, $\tilde{\sigma}=2 \sigma$, in terms of which the
field content (\ref{content}) reduces to 8 $\tilde{\sigma}$-periodic
bosons $X_I$ and chiral fermions $S_a$ and 8 $\tilde{\sigma}$-antiperiodic
fermions $S_{\dot{a}}$ with all possible values of $P_{\tilde{\sigma}}$
momenta. The partition function (or elliptic genus, since we are in the
odd spin structure for the $S_a$ fields) is given by  
\be
\frac{8-8}{\tau_2^{4-d/2}}
\frac{\vartheta_4^4(\tilde{q})}{\eta^{12}(\tilde{q})}
\sum_{P\in \Gamma_{d,d}}{\tilde{q}}^{p^2/2}\bar{\tilde{q}}^{\bar{p}^2/2}  
\label{ZD16}
\ee
where right-moving oscillators cancel out between the $X_I$ and $S_a$
supersymmetric fields. As before, the factor $(8-8)$ takes into account the 
ground-state multiplicities of the short BPS supermultiplets 
and $\tilde{q}\equiv e^{2\pi i \tau}$.
The partition function (\ref{ZD16}) 
reproduces the correct masses, charges and degeneracies
coming from (\ref{theta4}) once the level matching condition (\ref{LM}) for
$N=1$ is implemented, by adding the $\tau\rightarrow \tau +1$ amplitude
(\ref{theta3}), and the radius of compactification of the dual theory
$\tilde{R}$ is identified with twice the radius $R$ of the original one.

We now proceed to study the $N>1$ case which, as previously 
stated, corresponds to $N$ copies of the $N=1$ field content
modded out by the permutation group ${\cal S}_N$. In this case we can use the
results of section 2 for the elliptic genus of symmetric 
products. In \cite{gmnt} it was argued that 
only the longest string sector $[g]=(N)$ represents a truly one-particle  
bound state of $N$ D-strings, while $(L)^M$ can be interpreted as an $M$-string
state, each of which represents a threshold bound state of $L$ strings. 
The contribution of the relevant sector ($M=1$, $L=N$)
can be read off from (\ref{genusIIB}), with
$\chi(\tilde{q})\,\bar{\chi}(\bar{\tilde{q}})$ given by (\ref{ZD16}), to be
\be
\frac{8-8}{\tau_2^{4-d/2}}
\sum_{s=0}^{N-1}\frac{\vartheta_4^4({\tilde{q}}^{\frac{1}{N}}\omega^s)}
{\eta^{12}(\tilde{q}^{\frac{1}{N}}\omega^s)}
\sum_{P\in \Gamma_{d,d}}\omega^{s \frac{P^2}{2}}
{\tilde{q}}^{p^2/2}\bar{\tilde{q}}^{\bar{p}^2/2}  
\label{ZD16N}
\ee       
with $\omega=e^{-\frac{2 \pi i }{N}}$ 
The sum over $s$ projects on the 
modes that satisfy
\be
k=\frac{1}{N}(\frac{P^2}{2}+N_R)\in {\bf Z}
\label{lm2}
\ee 
which is just the level matching condition (\ref{LM}). The degeneracies
can be found upon expanding (\ref{ZD16N}) in powers of 
$\tilde{q}^{\frac{1}{N}}$.
In particular for $N$ odd they come from the expansion of 
$\vartheta_4(\tilde{q}^{\frac{1}{N}})$ reproducing the degeneracies arising
from the sum of (\ref{theta4}) and (\ref{theta3}) 
once we appply the level matching
condition (\ref{lm2}). For $N$ and  $s$ both even, additional fermionic 
zero-modes that we should omit in the sum (\ref{genusIIB}) appear for the 
$S_{\dot{a}}$ field with $\alpha_\phi=\frac{1}{2}$ and $\beta_\phi=t_g=t_h=0$ 
in (\ref{modbc}). For the remaining
values of $s$ we find $\vartheta_2({\tilde{q}}^{\frac{1}{N}}
\omega^s)$ $(\alpha=\frac{1}{2}$, $\beta=0)$, which reproduces the
degeneracies, masses and charges coming from (\ref{theta2}). 

\subsection{ Type IIB on $(K3\times S^1)/(-)^{F_L}\sigma_V$}  

Finally, we apply our previous analysis to a model
with a lesser number of supersymmetries. We take as an example the
type II $(2,0)$ model arising from the orbifolding of 
$(K3\times S^1)$ by $(-)^{F_L}\sigma_V$, where $\sigma_V$ is a shift
of order two on the circle $S^1$. As before left-moving 
supercharges are projected out by the orbifold group, 
while the $\sigma_V$ shift ensures
that no new supersymmetries appears from the twisted sector.
In the effective five-dimensional theory only two gauge fields
are left out by the projection: the K-K reduction of the metric
$G_{\mu 5}$ and NS-NS antisymmetric tensor $B^{^{NS}}_{\mu 5}$.

For semplicity, we will discuss in detail the orbifold limit $T^4/I_4$ 
of $K3$ with $I_4$ the reflection of the four torus coordinates. 
The elliptic genus encoding masses, multiplicities and charges for
the corresponding BPS perturbative spectrum is then given by
%
%
\begin{eqnarray}
Z_{JI}&=&-\frac{(4-4)_{V}}{\tau_2^{4-d/2}}
         \frac{\vartheta_3(q)^2\vartheta_4^2(q)}{\eta(q)^6\vartheta_2^2(q)}
        \sum_{P\in \Gamma_{1,1}+V}q^{p^2/2} 
        \bar{q}^{\bar{p}^2/2}\label{34}\\
Z_{KI}&=&-\frac{(4-4)_{H}}{\tau_2^{4-d/2}}
         \frac{\vartheta_4^2(q)\vartheta_2^2(q)}{\eta(q)^6\vartheta_3^2(q)}
        \sum_{P\in \Gamma_{1,1}+V}q^{p^2/2} 
        \bar{q}^{\bar{p}^2/2}\label{24}\\
Z_{KJ}&=&-\frac{(4-4)_{H}}{\tau_2^{4-d/2}}
         \frac{\vartheta_3^2(q)\vartheta_2^2(q)}{\eta(q)^6\vartheta_4^2(q)}
        \sum_{P\in \Gamma_{1,1}+V} 
        e^{2\pi i V\cdot P}
         q^{p^2/2}\bar{q}^{\bar{p}^2/2}\label{23}\\
Z_{IK}&=&\frac{(4-4)_{H}}{\tau_2^{4-d/2}}
         \frac{\vartheta_4^2(q)\vartheta_2^2(q)}{\eta(q)^6\vartheta_3^2(q)}
        \sum_{P\in \Gamma_{1,1}} 
        e^{2\pi i V\cdot P}
        q^{p^2/2}\bar{q}^{\bar{p}^2/2}\label{24e}\\
Z_{IJ}&=&\frac{(4-4)_{H}}{\tau_2^{4-d/2}}
         \frac{\vartheta_3^2(q)\vartheta_2^2(q)}{\eta(q)^6\vartheta_4^2(q)}
        \sum_{P\in \Gamma_{1,1}} 
        e^{2\pi i V\cdot P}q^{p^2/2}\bar{q}^{\bar{p}^2/2}
\label{23e}
\end{eqnarray}
The subscripts refer as usual to the different twists 
in the $\sigma$ and $\tau$ direction respectively, 
$J\equiv(-)^{F_L}\sigma_V$, $I\equiv I_4$ and $K=I\cdot J$.
$\Gamma_{1,1}$ denotes the even self-dual Lorentzian lattice of 
generalized momenta and $V=(v,\bar{v})= ({R}/{2},-{R}/{2})$. 
The sectors $Z_{IJ}$ and $Z_{KJ}$ are the T-modular transforms
($\tau\rightarrow \tau+1$) of $Z_{IK}$ and $Z_{KI}$ respectively,
ensuring the level matching condition. The $(4-4)$'s factors
with subscripts $V$ and $H$ are associated to the 4 bosons and fermions
of the vector and hyper five-dimensional short ${\cal N}=4$ 
supermultiplets\footnote{We are still
adhering to the four-dimensional counting of supersymmetries.} respectively. 
Notice  that the above model allows the introduction of discrete torsion
\cite{distor}, \ie an opposite sign in the $J$ projection of the $I$ sector.
This would result in a reapparance of R-R massless states in the $K$ sector.
The same freedom is possible in the dual type I model, where it corresponds
to retaining the (descendants of the 6D) tensor multiplets instead
of the hypermultiplets in the twisted sector of the orbifold.
Notice that the presence of R-R massless states in type II (2,0) models
opens new problems in the duality with type I models without open strings
that we will not address here.

For this reason we will restrict our attention to the dual version of the 
type II $(2,0)$ model discussed  above. The type I dual can be constructed as
before mapping the $(-)^{F_L}$ operation to
the $\Omega$-projection under type IIB S-duality \cite{sen}. 
Notice that the combination
of $\Omega$ and $\sigma_V$ leads to a rather unconventional type I vacuum
configuration in which no D9-branes and their open-string excitations can be
introduced. More explicitly the Klein-bottle amplitude 
\be
{\cal K} = Tr_c (\Omega \cdot \sigma_V q^H)
\label{unklein}
\ee
where the subscript $c$ denotes a trace in the unoriented closed-string spectrum
does not give rise to {\it unphysical} massless tadpoles in the transverse
channel. Indeed no massless state at all participate to the
crosscap-to-crosscap amplitude. After target space T-duality along the circle
$S^1$, this may be
interpreted as saying that there are two oppositely charged O8-planes
\cite{wittor}. 
The overall vanishing of the R-R charge prevents the introduction
of D-branes and their associated open-string amplitudes (annulus ${\cal A}$ and
M\"obius-strip ${\cal M}$) \cite{dp,gep,mbtor}. In some respect
type I vacuum configurations without open strings may be considered as 
unconventional orientifolds without twisted sectors \cite{ascar,hor,bs}.
The presence of the $K3$ factor in the compactification manifold does not
change the picture qualitatively with respect to the previous subsection. 
We are left with the
same field content (\ref{content}) as in the previous (toroidal) case 
now with target space
$(R^4\times K3)^N/{\cal S}_N$. For $N=1$ the partition function can be read
off directly from the orbifolding of (\ref{ZD16}) 
\bea
Z_{+-}&=&\frac{({4}-{4})}{\tau_2^{4-d/2}}
\frac{\vartheta_3^2(\tilde{q})\vartheta_4^2(\tilde{q})}
{\eta^{6}(\tilde{q})\vartheta_2^2(\tilde{q})}\label{34'}\\
Z_{--}&=&\frac{({4}-{4})}{\tau_2^{4-d/2}}
\frac{\vartheta_2^2(\tilde{q})\vartheta_4^2(\tilde{q})}
{\eta^{6}(\tilde{q})
\vartheta_3^2(\tilde{q})}\label{24'}
\eea
where now $\pm$ refer to the $I_4$ twists. The spectrum of type I
D-strings accounted for by (\ref{34'}) and (\ref{24'}) agrees with the
spectrum of fundamental strings accounted for by (\ref{34}) and (\ref{24}) once
the level matching condition is imposed. The remaining states (\ref{24e}) will
appear for $N$ even. Indeed extracting the longest string contribution
from expression (\ref{chiab}) for $N$ odd, we recognize the same partition
function (\ref{34'}-\ref{24'}) as before after the substitution $q\rightarrow
q^{\frac{1}{N}}\omega^s$ and the sum over $s=0,\dots ,N-1$ that 
ensures the level matching condition (\ref{lm2}). 
For $N$ even, on the other hand, we find
\be
Z_{--}=\frac{(\bar{4}-\bar{4})}{\tau_2^{4-d/2}}
\sum_{s=0}^{N-1}\frac{\vartheta_2^2(\tilde{q}^{\frac{1}{N}}\omega^s)
\vartheta_4^2(\tilde{q}^{\frac{1}{N}}\omega^s)}
{\eta^{6}(\tilde{q}^{\frac{1}{N}}\omega^s)
\vartheta_3^2(\tilde{q}^{\frac{1}{N}}\omega^s)}
\ee     
which agrees with the fundamental (\ref{24e}) spectrum of states.

\section{Threshold corrections in type II string vacua}

In the previous sections we provided some evidence for 
various equivalences between string theories by
studying the spectrum of physical BPS states in lower dimensional
compactifications. 
We would now like to apply these results to the study 
of the low energy effective actions describing these string vacua. 
In particular, we are interested in studying the
moduli dependence of the special (``BPS saturated'') ${F}^4$ terms in the
$D=8$ dimensional low energy effective action for the conjectured pair
type II (4,0) string - type I without open strings.
We closely follow a sequence of works 
\cite{bfkov}, where a similar analysis for
the threshold corrections in the context of type I - heterotic duality
has been performed. 

The interest in the study of these terms relies
on the fact that they are believed to receive only one-loop corrections
for toroidal
heterotic compactifications to $D>4$ dimensions. 
Supersymmetry protects these terms from higher-loop perturbative corrections
while the only identifiable
source of non-perturbative corrections (the 5-brane instanton) 
is infinitely heavy for compactifications to $D>4$ dimensions.
This seems to be the case for the type II (4,0) models too. 
For the model under consideration, the $(-)^{F_L}\sigma_V$ action removes 
the RR-fields leading to an effective type II theory without D-branes
\cite{dp,gep,mbtor}.
The only source of non-perturbative effects we can think of is again 
the 5-brane instanton which cannot enter the correction of a $D>4$   
dimensional effective action. 
On the type I side (type I without
open strings or type IIB on $T^2/\Omega\sigma_V$ \cite{mbtor}) however, 
D-string instantons are expected to correct the effective actions for $D\leq 8$
dimensions. Indeed, using the conformal description of the infrared limit
for the $N$ 
D-instanton\footnote{D-instantons in this context refer to instanton from
the point of view of the eigth-dimensional effective action. The
``D" recalls the origin of these contributions from D-strings
wrapped on the $T^2$ torus.} system we will be able to compute these 
non-perturbative corrections, showing the agreement with the one-loop
exact formula found in the dual type II computation.   

\subsection{Threshold corrections in type IIB on 
$T^2/(-)^{F_L}\sigma_V$}
 
For simplicity we will restrict our attention 
to the ${F}^4$ couplings in the low energy effective action. 
We will assume that, as it is the case for similar terms in toroidal 
heterotic compactifications to $D>4$ dimensions, the moduli
dependence for these terms in type IIB on $T^2/(-)^{F_L}\sigma_V$
receive only one-loop corrections. 
Let us recall how the arguments
leading to this conclusion for the $SO(32)$ heterotic ${F}^4$
terms work. The ${F}^4$ terms we will study 
in the type II context involve the $O(2,2)$ gauge fields 
$G_{\mu i}, B_{\mu i}$ arising
from the K-K reduction of the metric and NS-NS
antisymmetric tensor. In toroidal compactifications of the 
heterotic string these gauge fields get mixed with the $SO(32)$ gauge fields
by the action of the T-duality group $O(2,18,{\bf Z})$.
It is therefore reasonable to believe that the same non-renormalization
arguments apply to all the gauge fields that do not belong to the supergravity
multiplet. As we will see, the latter, usually called ``graviphotons", stay on 
a different footing. For $SO(32)$ gauge fields,
the ${F}^4$ terms (as well as some ${\cal R}^4$ and 
${F}^2{\cal R}^2$ terms) can be obtained
by dimensional reduction of ten dimensional superinvariants,
whose bosonic parts read \cite{rooTseytlin}
\bea
I_1 &=& t_8tr{F}^4-\frac{1}{4}\varepsilon_{10} B tr{F}^4
\nonumber\\ 
I_2 &=& t_8(tr{F}^2)^2-\frac{1}{4}\varepsilon_{10}B(tr{F}^2)^2
\label{invariants}
\eea
They are special because they contain CP-odd pieces related to 
the cancellation of gravitational and gauge anomalies in ten dimensions. 
Indeed in ten dimensions, the coefficients of these couplings are 
completely determined by supersymmetry and the 
anomaly cancelling mechanism.
This is no longer true for compactifications to lower dimensions, where
supersymmetry restricts, but does not completely fix, their dependence on the
compactification moduli. For heterotic compactifications,
the moduli dependence of the CP-odd pieces in (\ref{invariants})
was studied in \cite{yasuda}. As shown there, they receive
only one-loop perturbative corrections. Moreover, as 
argued before, non-perturbative corrections
are ruled out in compactifications to $D>4$ dimensions. 
It is then plausible to assume that no higher-order  
corrections are present for the supersymmetry-related 
CP-even ${F}^4$ terms, too. 

A similar analysis for the type II models under study here, 
has not been done, but we expect similar result to be true.
We will assume that this is the case, \ie 
that the one-loop formula we will obtain in this section for
the moduli dependence of some ${F}^4$ terms in type IIB on 
$T^2/(-)^{F_L}\sigma_V$ is exact. The non-perturbative
results for similar terms in the low energy effective action 
of the dual type I string will support this assumption. 

We will consider a compactification on a target space torus characterized
by the complex moduli
\bea
 T &=& T_1 + iT_2 = \frac{1}{\alpha' } (B^{^{NS}}_{89} + i\sqrt{G})\nonumber\\
 U &=& U_1 + i U_2 = ( G_{89} + i\sqrt{G} )/ G_{88} \ ,
\eea
where $G_{ij}$ and $B^{^{NS}}_{ij}$  are the
$\sigma$-model metric and NS-NS antisymmetric
tensor. 

We recall that by projecting the type IIB string with $(-)^{F_L}\sigma_V$ we
remove all R-R massless fields. The only eight-dimensional gauge bosons     
are then given by the $G_{\mu8},G_{\mu9}$ 
and $B^{^{NS}}_{\mu8},B^{^{NS}}_{\mu9}$ components of the metric and 
NS-NS antisymmetric
tensor. The corresponding vertex operators in the Green-Schwarz 
formalism can be written as  
\bea
V^R_i &=& \int d^2 z (G_{\mu i}-B^{^{NS}}_{\mu i})(\partial X^{i}-\frac{1}{4} 
p_{\nu} S\gamma^{i \nu}S)(\bar{\partial} X^{\mu}-\frac{1}{4} 
p_{\rho} \tilde{S}\gamma^{\mu \rho}\tilde{S})e^{ip X}
\label{graviphotonvertex}\\
V^L_i &=& \int d^2 z (G_{\mu i}+B^{^{NS}}_{\mu i})(\partial X^{\mu}-\frac{1}{4} 
p_{\nu} S\gamma^{\mu \nu}S)(\bar{\partial} X^{i}-\frac{1}{4} 
p_{\rho} \tilde{S}\gamma^{i \rho}\tilde{S})\label{gaugevertex}  
e^{i p X}  
\eea 
with $i=8,9$. The $V^R_i$ are the vertices for the graviphotons, \ie the
gauge fields sitting in the supergravity multiplet. It is easy to see
that there are no one-loop corrections to ${F}^4$ terms involving the 
graviphotons. 
Indeed, after soaking the eight right moving zero-modes $S_0$ on the world-sheet
torus with the fermionic piece $p_{\nu}S\gamma^{i\nu}S$ in 
(\ref{graviphotonvertex}), a four-graviphoton amplitude already exposes
the required fourth power of the external momenta. At this order in 
the momenta, only the bosonic piece in the left-moving parts of the
vertices can enter, but their contractions unavoidably lead to total 
derivatives
$\langle\bar{\partial}X^{\mu}(z_1)\bar{\partial}X^{\mu}(z_2)\rangle$ 
which vanish after the $z$ -integrations.
Corrections to the ${{F}^4}$ terms for the $G_{\mu i}$ and  
$B^{^{NS}}_{\mu i}$ gauge fields then coincide and can be extracted from
the four-point amplitude  
\bea
{\cal A}_{\ell}=\langle (V^L_8)^\ell (V^L_9)^{4-\ell} \rangle=
t_8 F_8^\ell F_9^{4-\ell}
\langle\prod_{j=1}^{\ell} \int dz_j\bar{\partial} X^8(z_j)
\prod_{k=\ell +1}^{4}\int dz_k\bar{\partial} X^9(z_k)\rangle 
\label{Aa}
\eea      
where $t_8$ is the tensor arising from the trace over the
right-moving fermionic zero-modes.   
It is convenient to define a generating function 
for such terms that reads
\be
{\cal A}_\ell = 
t_8 F_8^\ell F_9^{4-\ell}\int_{\cal F} \frac{d^2 \tau}{\tau_2^2}
(\frac{\tau_2}{\pi})^4 \frac{\partial^\ell}{\partial\nu_8^\ell}
\frac{\partial^{4-\ell}}{\partial\nu_9^{4-\ell}}Z(\nu_i,\tau,\bar\tau)
\label{ampl}
\ee
with ${\cal F}$ the fundamental domain for the world-sheet torus, and 
$Z(\nu_i,\tau,\bar\tau)$ the partition function arising from
a perturbed Polyakov action whose bosonic part reads 
\be
S({\nu_i}) = \frac{2\pi}{\alpha^\prime} \int d^2 \sigma ( \sqrt{g}
  G_{\mu\nu}g^{\alpha\beta}\partial_\alpha X^\mu \partial_\beta X^\nu
+i  B^{^{NS}}_{\mu\nu}\varepsilon^{\alpha\beta}
\partial_\alpha X^\mu \partial_\beta X^\nu+
\sqrt{g}\frac{\alpha^\prime}{2\tau_2}\nu_i\bar{\partial}X^i ).
\label{polyakov}
\ee              
and
$\bar{\partial}=
\frac{1}{\tau_2}(\partial_{\sigma_2}-\bar{\tau}\partial_{\sigma_1})$.
The partition function $Z(\nu_i,\tau,\bar\tau)$ 
involves a sum over all possible world-sheet instantons
\bea
\pmatrix{X^8\cr X^9}=
 M \pmatrix{\sigma^1\cr \sigma^2}\equiv
\pmatrix{m_1&n_1 \cr
m_2&n_2}\pmatrix{\sigma^1 \cr \sigma^2}
\label{M}
\eea
with worldsheet and target space coordinates $\sigma_1,\sigma_2$ and 
$X^8,X^9$ respectively, both taking values in the interval (0,1]. 
The entries $m_1,n_1$ are integer or
half-integer depending on the specific orbifold sector, while $m_2,n_2$
are always integers. We denote the three relevant sectors: $n_1$ half-integers,
$m_1$ half-integers and both half integers by $\epsilon=+-,-+,--$ respectively.
Clearly the untwisted sector, $\epsilon=++$, will not contribute since it has
too many zero modes to be soaked by the four vertex insertions at this order
in the momenta. 

In the following we use normalizations such that
$\langle X^{M} X^{N} \rangle\sim G^{M N}$,
with $G^{M N}$ the ten dimensional
metric defined by 
\be
G_{i j} = \frac{\alpha^\prime T_2}{U_2}\, 
\pmatrix{ 1 & U_1 \cr U_1 & | U |^2 \cr}
\label{metricij}
\ee
in the compact space and the flat metric $G_{\mu\nu}=\eta_{\mu\nu}$ in 
${\bf R}^8$.
For the worldsheet metric we choose
\be
g^{\alpha\beta} = \frac{1}{\tau_2^2}
\,  \pmatrix{ |\tau|^2 & -\tau_1 \cr
-\tau_1 & 1 \cr}\ . 
\ee
The generating function can then be written as
\be
\int_{\cal F}\frac{d^2 \tau}{\tau_2^2} Z(\nu_i,\tau\bar\tau) =
\frac{{\cal V}_8}{2^{10}\pi^4}\int_{\cal F}
\frac{d^2 \tau}{\tau_2^2} 
\sum_{\epsilon}\Gamma^{\epsilon}_{2,2}(\nu_i){\cal A}_\epsilon
\label{Zl}
\ee
with 
\be
\Gamma^{\epsilon}_{2,2}(\nu_i)=
\frac{T_2}{U}\,
 \sum_{M_{\epsilon}}
 e^{ 2\pi i  T {\rm det}M }
e^{- \frac{\pi T_2 }{ \tau_2 U_2 }
\big| (1\; U)M  \big( {\tau \atop -1} \big) \big| ^2}
e^{- \frac{\pi}{ \tau_2}
(\nu_8\,\nu_9)M  \big( {\tau \atop -1} \big) }
\label{lattice}
\ee
and ${\cal A}_\epsilon$ the anti-holomorphic BPS partition functions 
(\ref{theta2}-\ref{theta3})
\bea
{\cal A}_{+-}\equiv
\frac{1}{2}\frac{\vartheta_2(\bar{q})^4}{\eta(\bar{q})^{12}}
\quad , \quad
{\cal A}_{-+}\equiv
\frac{1}{2}\frac{\vartheta_4(\bar{q})^4}{\eta(\bar{q})^{12}}
\quad , \quad
{\cal A}_{--}\equiv
\frac{1}{2}\frac{\vartheta_3(\bar{q})^4}{\eta(\bar{q})^{12}}
\label{aepsilon}
\eea

Following Dixon, Kaplunovsky and Louis \cite{dklo} we can express the
sum over the $M_{\epsilon}$ matrices in (\ref{lattice}) as a sum over 
$SL(2,Z)$ representative integrated in an unfolded domain.     
Notice that the complete generating function $Z(\nu_i,\tau,\bar\tau)$ has 
no definite modular transformation properties. This is not the case
for the interesting term, the fourth $\nu$-derivative 
of $Z(\nu_i,\tau\bar\tau)$  
appearing in (\ref{Aa}), which is indeed modular invariant. In the
following we keep in mind that eventually we will only consider 
this term and, without further comments, perform modular manipulations 
which are only sensible on the final result.    
In this broad sense the generating function (\ref{Zl}) is invariant
under the combined $SL(2,Z)$ actions
\be
\tau \rightarrow \frac{a\tau+b}{c\tau +d}\ , \ \ {\rm and}\ \ 
M_\epsilon \rightarrow  
M_\epsilon \ \left( \matrix{d & b\cr c & a\cr}\right) \ 
\label{transf}
\ee
Different $\epsilon$-elements in the sum (\ref{lattice}) get mixed 
in general by these 
transformations, but the sum is clearly invariant. 
An orbit is defined by the set of matrices $M$, which can be
related by some $SL(2,Z)$ element $V$ to a given representive 
$M_0$ through $M=M_0 V$.
By a change of variables in the $\tau$ integration we can reduce the 
sum over matrices $M$ in a given orbit to a single integration over
an unfolded domain obtained as the union of $V_i$ images of the 
fundamental domain through the modular transformations (\ref{transf}).   
These unfolded domain can be either the strip or
the whole upper half plane depending on whether the matrix M is 
{\it degenerate} (det $M=0$) or {\it non-degenerate} (det $M\neq 0$).

Let us consider first the degenerate case. Using the modular transformation
properties of (\ref{lattice}) and (\ref{aepsilon})
we can write the contributions from the different
orbifold sectors in (\ref{Zl}) as
\bea
\int_{\cal F}\frac{d^2 \tau}{\tau_2^2} 
(\Gamma^{+-}_{2,2}{\cal A}_{+-}\,(\bar\tau)+
\Gamma^{+-}_{2,2}{\cal A}_{+-}\,(\bar\tau')+
\Gamma^{+-}_{2,2}{\cal A}_{+-}\,(\bar\tau''))
=\int_{{\cal F}_2}\frac{d^2 \tau}{\tau_2^2} 
\Gamma^{+-}_{2,2}{\cal A}_{+-}(\tau)
\label{deg+-}
\eea
where $\tau'=-\frac{1}{\tau}$ and $\tau''=\tau'+1$. The new fundamental
domain ${\cal F}_2$ is the quotient 
of the upper half plane by a 
$\Gamma_2$ subgroup of $SL(2,Z)$ transformations defined as the
set of elements $V$ which keep invariant the form of an $M_{+-}$ matrix 
($m_1\in {\bf Z},n_1\in {\bf Z}+\frac{1}{2}$). 
A generic element of $\Gamma_2$ can be written as 
\be
V= \left( \matrix{a  & b\cr 2 c'  & d \cr}\right).
\label{Vdeg}
\ee  
with $ad-2bc'=1$. It is easy to see that acting with a $V$ in
this $\Gamma_2$ subgroup  
we can always bring a matrix $M_{+-}$ with zero determinant 
to the form
\be
M  = \left( \matrix{0  & j_1-\frac{1}{2}\cr 0  & j_2 \cr}\right).
\label{Mdeg}
\ee    
The sum over $M_{+-}$ matrices in (\ref{deg+-}) can then be written as  
a sum over the representatives (\ref{Mdeg}) labeled by $(j_1,j_2)$.
The $\tau$ integration runs over an unfolded domain defined 
by the union of all images of the $F_2$ fundamental 
domains under the $\Gamma_2$ actions (\ref{Vdeg}). We should notice
however that not all $\Gamma_2$ matrices define different $M$'s.
Indeed, a representative (\ref{Mdeg}) is invariant under the action 
\be
V= \left( \matrix{1  & b\cr 0  & 1 \cr}\right)~~~{\rm i.e.} 
~~\tau\rightarrow \tau+b.
\label{Vtt1}
\ee   
The unfolded domain is then the upper half plane modded out by these
transformations, \ie the strip ${\cal S} = \{|\tau_1|\leq \frac{1}{2}, 
\tau_2>0 \}$. 
Substituing (\ref{Mdeg}) in (\ref{deg+-}) we are left with 
\be
\int_{\cal S} {d^2\tau\over\tau_2^2}
\sum_{(j_1,j_2) \neq (0,0)}
 e^{- \frac{\pi T_2 }{ \tau_2 U_2 }
\big| j_1-\frac{1}{2}+j_2U \big| ^2 } 
e^{-\frac{\pi }{ \tau_2}(\nu_8(j_1-\frac{1}{2})+\nu_9 n_2)}
\,{\cal A}_{+-}(\bar{\tau}).
\ee
The $\tau_1$ integration picks the $\bar{q}^0$ power in the expansion
\be
{\cal A}_{+-}(\bar{\tau}) =\sum_{n=0}^\infty {\cal A}^n_{+-} \bar{q}^n.
\label{aqn}
\ee
which is just $2^3$. After taking the $\nu$-derivatives and
integrating in $\tau_2$ we are left with the final result
\bea
&&\langle (F_8)^\ell (F_9)^{4-\ell} \rangle_{deg}=\nonumber\\
&&\frac{{\cal V}_8}{(4\pi^2\alpha^\prime)^4}t_8 F_8^\ell F_9^{4-\ell}\,T_2 
\int_0^\infty {d\tau_2\over\tau_2^6}\,
\sum_{(j_1,j_2) \neq (0,0)}(j_1-\frac{1}{2})^\ell j_2^{4-\ell}
 e^{- \frac{\pi T_2 }{ \tau_2 U_2 }\big| j_1-\frac{1}{2}+j_2U \big| ^2 }
\label{ffd}
\eea
Let us now consider the contributions from the non-degenerate orbits.  
In this case acting with an $SL(2,Z)$ transformation we can, at most,
bring a matrix $M$ to the form
\be
M  = \left( \matrix{m_1  & n_1\cr 0  & n_2 \cr}\right)
\label{Mnondeg}
\ee  
where $m_1,n_1$ are in $\bf Z$ or ${\bf Z}+\frac{1}{2}$ 
as before depending on the
orbifold sector. The $\tau\rightarrow \tau+b$ action (\ref{Vtt1}) on this
matrix shifts $n_1\rightarrow n_1+b\, m_1$. Therefore we can bring 
$n_1$ to the fundamental range
$n_1=0(\frac{1}{2}),1(\frac{3}{2}),\cdots, 0(\frac{1}{2})+m_1-1$. 
The representatives
of the non-degenerate matrices $M$ are then given by the matrices 
(\ref{Mnondeg}) with
\bea
\epsilon&=&+- \quad : \quad ~~~m_1\in {\bf Z};
~~~~~~~~n_1=\frac{1}{2},\frac{3}{2},\cdots,\frac{2 m_1-1}{2}
\nonumber\\   
\epsilon&=&-+ \quad : \quad ~~~m_1\in {\bf Z}+\frac{1}{2};~~~
n_1=0,1,\cdots m_1-1
\nonumber\\   
\epsilon&=&-- \quad : \quad ~~~m_1\in {\bf Z}+\frac{1}{2};
~~~n_1=\frac{1}{2},\frac{3}{2},\cdots,\frac{2 m_1-1}{2}
\label{mnnondeg}
\eea
The unfolded domain is now the whole upper half complex plane since all
$SL(2,Z)$ actions are allowed, \ie distinct elements in a non-degenerate
orbit are in one-to-one correspondence with the copies of the
fundamental domain ${\cal F}$
of $\tau$ in the upper half plane.
We should however notice that different replicas of the 
fundamental domain in the upper half plane come from different
orbifold sectors since   
$SL(2,Z)$ transformations which allow us to bring a given matrix
$M$ to a representative in (\ref{mnnondeg}) mixed the 
$M_{\epsilon}$ with different $\epsilon$'s.
Therefore only the unfolding of the sum over $\epsilon$'s
in (\ref{Zl}) makes sense.
  
We can now perform the $\tau$ integrations. Expanding the antiholomorphic
modular functions (\ref{aepsilon}) as in (\ref{aqn}) yields the
integral
\be
I_{n} = 
T_2 e^{2\pi i T m_1 n_2}\int_{{\bf C}^+} \frac{d^2\tau}{\tau_2^{2}}
         \, e^{- \frac{\pi T_2 }{ \tau_2 U_2 }
        \big|m_1\tau - n_1-n_2 U \big| ^2 }
   e^{- \frac{\pi}{ \tau_2 }
        [(m_1\tau - n_1)\nu_8-n_2 \nu_9] } e^{-2i\pi\bar{\tau} n}
\label{In}
\ee
which after the $\tau$ integrations can be written as
\bea
I_{n}&=&\frac{(U_2 T_2)^{\frac{1}{2}}}{m_1}
e^{-2\pi i T_1 m_1 n_2} 
e^{-2\pi i n(\frac{n_1+U_1 n_2}{m_1})}
e^{\pi i \nu_8(\frac{n U_2}{m_1 T_2}-m_1)}
\int_0^\infty
\frac{d\tau_2}{\tau_2^{3/2}} e^{-\frac{\beta}{\tau_2}-\gamma\tau_2}\nonumber\\
&=&\frac{(U_2 T_2)^{\frac{1}{2}}}{m_1}
e^{-2\pi i T_1 m_1 n_2} 
e^{-2\pi i n(\frac{n_1+U_1 n_2}{m_1})}
e^{\pi i \nu_8(\frac{n U_2}{m_1 T_2}-m_1)}
\sqrt{\frac{\pi}{\beta}}e^{-2\sqrt{\beta\gamma}}
\label{Inf}
\eea
with 
\bea
\beta&=& \pi (n_2^2 U_2 T_2+\nu_9 n_2 -\nu_8 U_1 n_2
-\frac{\nu_8^2 U_2}{4 T_2})\nonumber\\
\gamma&=&\frac{\pi T_2}{U_2}(m_1+\frac{n U_2}{T_2 m_1})^2.
\eea

In order to extract the ${F}^4$ term we should
still act on (\ref{Inf}) with four $\nu$-derivatives 
and finally set the sources $\nu$'s to zero. 
In the following we will restrict ourself to the leading 
behaviour in a $\frac{1}{T_2}$ expansion of this result. The interest
in this particular expansion will become clear later. In this
limit the four derivatives should hit one of the $\nu$-linear terms in
the exponential of (\ref{Inf}) leaving the final expression
\bea
&&\langle (F_8)^\ell (F_9)^{4-\ell} \rangle_{nondeg}=\nonumber\\
&&=\frac{{\cal V}_8}{(4\pi^2\alpha^\prime)^4}t_8 F_8^\ell F_9^{4-\ell}
\sum_{n,\epsilon}\sum_{m_1,n_1,n_2}{\cal A}^n_\epsilon 
e^{-2 \pi i n(\frac{n_1+\bar{U}n_2}{m_1})}e^{-2\pi i m_1 n_2 T}
\frac{\partial^\ell}{\partial\nu_8^\ell}
\frac{\partial^{4-\ell}}{\partial\nu_9^{4-\ell}}
e^{\frac{\pi m_1}{U_2}(\nu_8 U-\nu_9)}\nonumber\\
&&=\frac{{\cal V}_8}{(4\pi^2\alpha^\prime)^4}t_8 F_8^\ell F_9^{4-\ell} 
\sum_{\epsilon}\sum_{m_1,n_1,n_2}\frac{m_1^4}{m_1|n_2|}\frac{U^\ell}{U_2^4}
e^{-2\pi i T m_1 |n_2|}{\cal A}(\frac{n_1+\bar{U}n_2}{m_1})
\label{ffn}
\eea  
where we have used the expansions (\ref{aqn}) to recontruct
the modular forms (\ref{aepsilon}), now evaluated at an induced modulus
${\cal U} = \frac{n_1+\bar{U} n_2}{m_1}$. 

\section{Threshold corrections in type I without open strings}

We now pass to the study of ${F}^4$ threshold corrections
in the conjectured dual (type IIB on $T^2/\Omega\sigma_V$) of 
the previously
studied type II orbifold model. If the one-loop formulas for
the moduli dependence of ${F}^4$ terms in type IIB on 
$T^2/(-)^{F_L}\sigma_V$ are exact, as argued before, they should contain
both the perturbative and non-perturbative corrections in the dual side.
The aim of this section is to prove that this is the case. The results are
in complete agreement with the predictions of the type IIB 
self-duality conjecture.    

Let us begin by translating the one-loop exact results 
(\ref{ffd}) and (\ref{ffn})
in terms of the type I variables. The duality relations 
imply a  rescaling of the lengths (in the $\sigma$ model variables) 
according to
\be
L_F=\frac{L_{I}}{\sqrt{\lambda_{I}}}   
\ee
where the subscripts ``$F$'' and ``$I$'' are used to distinguish orbifold
($T^2/(-)^{F_L}\sigma_V$) and
orientifold ($T^2/\Omega\sigma_V$) compactifications of the type IIB string.
This implies in particular that the 
volume $T_2$ gets rescaled as $T_2^F=T_2^{I}/\lambda_{I}$ and
therefore the expansion in $1/T_2$ of the exact result found in the previous
section can be identified with the genus expansion as seen from the type I 
perspective.
Taking into account also the scaling of the gauge field 
$A^F_\mu=G^F_{\mu i}=G^I_{\mu i}/\lambda_I=A^I_\mu/\lambda_I$,
we find that the relevant ${F}^4$ terms in the eight dimensional
effective action scales according to
\be
f(T_2^F,\lambda_F)\int d^8 x\sqrt{G_F}\,F_F^4=
\frac{1}{\lambda_I^4}\,f(\frac{T_2^I}{\lambda_I},\frac{1}{\lambda_I})
\int d^8x\sqrt{G_I}\,F_{I}^4 \quad .
\label{f4eff}
\ee
In the previous section we argued that the only non-trivial moduli
dependence for these terms in the orbifold side are given by the
one-loop ($\lambda_F^0$ order in the expansion of  $f(T^F_2,\lambda_F)$) 
expressions (\ref{ffd}) and (\ref{ffn}). By plugging these results in
(\ref{f4eff}) we can see that contributions from
degenerate orbits ($f(T_2^F)\sim (T^F_2)^{-4}$) correspond to
one-loop effects (order $\lambda_I^0$) in the type I description, while 
those from non-degenerate matrices 
($f(T^F_2)\sim e^{-2 \pi T^F_2 m_1 n_2}(1+O(1/T^F_2$)) should
arise as D-instanton corrections with instanton number $N=m_1 n_2$.       
We will momentarily show how these corrections can be 
reproduced by a direct computation in the type I theory.  

\subsection{One loop threshold corrections}  

The one-loop effective action for a type I theory without open
strings gets contribution from the torus and Klein bottle amplitudes.
We are interested in the corrections to ${F}^4$ terms. 
As before there are four eight-dimensional gauge fields 
$G_{\mu i}$ and $B^{^{RR}}_{\mu i}$, but only the former couple 
to elementary states (the K-K modes) in the type I string spectrum.  
As before the relevant vertex operators are defined by
\be
V_i= \int d^2z G_{\mu i}(\partial X^{\mu}-\frac{1}{4} 
p_{\nu} S\gamma^{\mu \nu}S)(\bar{\partial} X^{i}-\frac{1}{4} 
p_{\rho} \tilde{S}\gamma^{i \rho}\tilde{S})e^{i p X},
\label{vk}
\ee
since $\Omega\sigma_V$ is simply the worldsheet parity $\Omega$ when acting
on a massless state. In a four-point
amplitude, the sixteen fermionic zero-modes on the world-sheet torus,
if soaked, would produce eigth power of the external momenta. 
Therefore four-derivative terms only receive 
contribution from the Klein-bottle 
amplitude, in which left- and right-moving zero-modes are identified.   
The Klein-bottle partition function is defined by    
\bea
{\cal K}&=&-\frac{1}{2}\int_0^\infty \frac{dt}{t} {\rm Tr}\,\, 
\Omega e^{\pi i k_8}\, e^{-\pi t(k_{_N} k_{_M} G^{^{MN}}+M^2)}\nonumber\\  
&=&\frac{{\cal V}_8}{(4\pi^2\alpha^\prime)^4}(8-8)\int_0^\infty \frac{dt}{t^6}
\sum_{n_i}e^{-\frac{\pi}{t} (n_i + \epsilon_i/2) (n_j + \epsilon_j/2) G^{ij}}
\label{zk}
\eea
where $\epsilon_8=1/2$ and $\epsilon_9=0$ and the metric
$G^{ij}$ is the inverse of (\ref{metricij}). As always the factor 
$(8-8)$ comes from the trace over the fermionic zero-modes 
and $\frac{1}{t^4}$ from the momentum integration in the non-compact
directions. The second expression in (\ref{zk}) only involves a sum
over the classical configurations 
since quantum bosonic and fermionic contributions cancel 
out by supersymmetry.  
The sign $e^{\pi i k_8}$ defines the action of $\sigma_V$ 
on a given state of K-K momentum $k_8$ running in the loop.
We have performed a Poiss\'on resummation on the integers $k_8, k_9$
expressing this projection as a half-shift in the 
Lagrangian mode $n_8$.  Due to this shift in the
Lagrangian mode (winding) no massless
closed string state flow in the transverse channel. This implies in particular
that the O9-planes do not carry R-R charge and therefore 
there is no room for the introduction of D9-branes and their
open string string excitations
\cite{dp,gep,mbtor}. 

The insertion of four vertex operators (\ref{vk}) in (\ref{zk}) will soak the
eight left-right symmetric fermionic zero modes reproducing the correct
momentum structure $t_8 F_8^\ell F_9^{4-\ell}$.   
For the remaining part of the vertices only the bosonic zero modes
are relevant, 
leading to four K-K insertions in (\ref{zk}).
Putting all together we are left with the final expression
\bea
&&\langle (F_8)^\ell (F_9)^{4-\ell} \rangle_{one-loop}=\nonumber\\
&&=\frac{{\cal V}_8}{(4\pi^2\alpha^\prime)^4}t_8 F_8^\ell F_9^{4-\ell}T_2 
\int_0^\infty {dt\over t^6}\,
\sum_{(j_1,j_2) \neq (0,0)}(j_1-\frac{1}{2})^\ell j_2^{4-\ell}
 e^{- \frac{\pi T_2 }{2 U_2 t }\big| j_1-\frac{1}{2}+j_2 U \big| ^2 }
\eea 
which precisely reproduces the contribution (\ref{ffd}) of the
degenerate orbits in the dual type IIB model on $T^2/(-)^{F_L}\sigma_V$.  
   
\subsection{D-Instanton contributions}

Let us now consider non-perturbative corrections in the type I description.
In eight dimensions the only identifiable source of non-perturbative effects
in the present model are the contributions from D-instantons, arising
from wrapping the D-string worldsheet on the two-torus target-space. 
Since the insertion of four gauge vertices can soak up at most eight fermionic
zero modes only $\frac{1}{2}$-BPS D-instantons can contribute.
We can use the results of the previous section. There we studied the 
partition function for the $N$-wrapped D-string excitations, 
by going to the infrared limit where the theory flows to an
orbifold conformal theory. Indeed we can
read directly the BPS $N$ D-instanton partition 
function from (\ref{genusIIB}) once
the one-loop $\tau$-parameter is identified with the complex
structure  $U$ of the target torus. 
This summarizes the quantum contributions to the partition function
in the D-instanton background. 
In addition to this we should include the classical contribution arising from  
the $N$ D-instanton action for this model. The bosonic part of this action
coincides with the one for $N$ type IIB D-strings and can be written as
\be
S_{\rm D-inst}=
{2 \pi \over \alpha^\prime}\sum_{t=1}^N\int d^2\sigma 
(\sqrt{g}g^{\alpha\beta}\frac{1}{\lambda_I}G_{\mu\nu}
+i B^{^{RR}}_{\mu\nu}\varepsilon^{\alpha\beta})
D_\alpha X_t^\mu D_\beta X_t^\nu
\label{instaction}
\ee
where $t$ labels the $N$ Cartan directions of the unbroken $U(1)^N$ gauge
group and $D_\alpha$ represent the supersymmetric covariant derivatives,
which can be written in a complex basis as
\bea
D X_t^\mu &=&\partial X_t^\mu-\frac{1}{4}p_\nu 
S_t\gamma^{\mu\nu}S_t\nonumber\\
\bar{D} X_t^\mu &=&\bar{\partial} X_t^\mu-\frac{1}{4}p_\nu 
\tilde{S}_t\gamma^{\mu\nu}\tilde{S}_t \quad .\nonumber
\eea
The $X$'s are always in the static gauge 
$X_t^8=\sigma^1, X_t^9=\sigma^2$ and the world-sheet modular parameter 
$\tau$ is identified with the complex structure $U$ of the target-space. 
Computing (\ref{instaction}) for a background with only non-trivial components
along $G_{ij}$ and $B^{^{RR}}_{ij}$, 
we are simply left with $S_{D-inst}= 2\pi N T_I$ where $T_I$ is the ``dual'' 
complexified K\"ahler modulus 
\bea
T_{I}=T_1+i T_2=
\frac{1}{\alpha^\prime}(B_{89}^{^{RR}}+i\frac{\sqrt{G}}{\lambda_I}) \quad .
\eea

In order to study ${F}^4$ couplings for the 
$G_{\mu i}$ components of the metric, we can turn on a
background $\nu_i$ for this field and extract the coupling
from the fourth  $\nu$-derivative.
Notice that only the classical part of the D-instanton partition
function will be modified by these insertions since quantum
correlators are always given by total derivatives which drop out
after the $z$-integrations.     
We can identify in (\ref{instaction}) the relevant coupling as
\be
S=2 \pi T_I N + \frac{N \pi}{\alpha^\prime U_2}[ 
(G_{\mu 9}+ U G_{\mu 8})D_z X^\mu+
(G_{\mu 9}-\bar{U} G_{\mu 8})D_{\bar{z}}X^{\mu}]+ \cdots 
\label{coupling}
\ee
where $z=\sigma_1+U\sigma_2$ is the complex worldsheet coordinate.
Each $G_{\mu i}$ insertion should soak up two right-moving fermionic
zero modes. These fermionic modes only enter the term with $D_z$. 
Therefore the four $\nu$-derivatives 
always hit this term in (\ref{coupling}) bringing a power of $ N U/U_2$
for each $G_{\mu 8}$ insertion and a power of $N/U_2$ for each $G_{\mu9}$
insertion.  
Collecting the different pieces:
\begin{itemize}
\item{The classical contribution: $e^{2 \pi i T_I N}$}
\item{The quantum contributions from (\ref{genusIIB}) omitting the ubiquituous
(8-8) factor and replacing $\tau$ by $U$}
\item{The fermionic zero mode trace: $t_8 F_8^\ell F_9^{4-\ell}$} 
\item{A factor of $N U/U_2$ for each $G_{\mu8}$ insertion and of $N/U_2$
for each $G_{\mu9}$}  
\end{itemize} 
yields the final result for the D-instanton contributions 
\bea
&&\langle (F_8)^\ell (F_9)^{4-\ell} \rangle_{D-inst}=\nonumber\\
&&=\frac{{\cal V}_8}{(4\pi^2\alpha^\prime)^4}t_8 F_8^\ell F_9^{4-\ell}
\frac{U^\ell}{U^4}
\sum_{N,L,M} L^4 e^{2 \pi i T N}\frac{1}{N}
\sum_{s=0}^{L-1}
\frac{\vartheta{\frac{L}{2}+\frac{1}{2}\brack  \frac{s}{2}+\frac{1}{2}}
(q^{\frac{M}{L}}e^{2\pi i \frac{s}{L}})^4}
{\eta^{12}(q^{\frac{M}{L}}e^{2\pi i \frac{s}{L}})}
\eea
which precisely reproduces the contributions of the non-degenerate orbits 
(\ref{ffn}) after trivial identifications. 

\section{Conclusions and perspectives}

In this paper we have analysed in detail dual pairs of unconventional
models with 16 and 8 supercharges. The key ingredient in our discussion
has been the correct identification of the D-string effective actions in the
corresponding type I like models.
In section 3, we have derived the one-loop partition function for both 
the $SO(16)$ CHL model and for its candidate type I dual \cite{mbtor,wittor}. 
Moreover we have also computed
the elliptic genus for the effective $O(N)$ theory governing the dynamics of
$N$ D-strings in this unconventional type I toroidal compactification.
The perturbative type I BPS states were shown to be in one-to-one
correspondence with heterotic BPS states with zero winding.
The non-perturbative type I BPS states, identified with $N$ D-string
bound-states in the longest string sectors, were shown to be in one-to-one 
correspondence with the heterotic BPS states with $N$ units of winding. 
In section 6, the perfect agreement between the 
BPS spectra of a type II (4,0) model in $D=8$ and its candidate 
dual type I model without open strings \cite{mbtor} 
has been the key ingredient in showing the matching
between the thresholds corrections to some $F^4$ terms in the two 
descriptions. The precise matching of the BPS spectra and the
explicit computations performed in section 6 lead us to conclude
that BPS-saturated thresholds must coincide for the $SO(16)$ dual pair we have
discussed as well as for any dual pair that passes the BPS precision test.
Therefore, although we have not explicitely worked out the threshold
corrections for CHL models and the corresponding duals, we do not expect any
basic difficulties.

In this paper we have only analysed the leading term in the coupling constant
expansion around D-string instanton sectors and have found that they
agree with the results on the fundamental string side. However the
latter calculation also predicts subleading corrections around
the D-string instantons. Such corrections also exist in the standard
heterotic-type I dual pairs \cite{bfkov} and their origin on D-string
instanton side has not yet been analyzed. This problem is under investigation
and it appears that the usual $\sigma$-model expansion around D-string 
instantons can account for such corrections \cite{ghmn}.

The presence of stable non-BPS states may well play a role 
in the corrections to non BPS-saturated couplings that are present
both in theories with a large amount of supersymmetry and 
in theories with lower or no supersymmetry at all. Computing non BPS-saturated 
thresholds and finding precise agreement for some of them may help
putting dualities for theories with lower (or none) supersymmetry on a 
firmer ground. Extending the present analysis to other D-brane bound-states, 
\eg D5-branes, in order to compute threshold corrections in lower 
dimensions (\eg $D=4$) does not seem 
obvious at all and deserves a case-by-case analysis.

Another interesting direction of application of the results 
presented in this paper is the computation of the prepotential for theories 
with 16 supercharges in $D=8$.
As far as the dependence on the $O(2,2)$-moduli
is concerned the F-theory description \cite{ls} seems to be a very 
powerful competitor to our approach.

Let us conclude by sketching the steps needed in order to explicitly
compute some of the BPS-saturated thresholds in the $SO(16)$ heterotic - type
I dual pair. For the $F^4$ term involving the vector bosons
of the $SO(16)$ gauge group, associated to the world-sheet current algebra at 
level $k=2$, one has to extract the four-derivative term in the four-point 
amplitude on the torus with even spin structures of four massless vectors.
The contribution of the odd spin structure is associated to the anomaly 
cancelling term in $D=10$. The trick of the 
generating function used in the previous sections may be of great use in this 
respect. Supersymmetry considerations lead us to conclude that the relevant
threshold gets only a one-loop contribution on the CHL heterotic side.
 
At the one-loop order, on the type I side, these terms only get contribution 
from the annulus ${\cal A}$
and M\"obius-strip ${\cal M}$ amplitudes. It should be easy to realize that
the contribution of the degenerate orbits to the heterotic threshold matches
the contribution of the perturbative type I amplitudes ${\cal A}$ and 
${\cal M}$. A somewhat involved computation may be required in order to match
the contribution of the non-degenerate orbits to the heterotic thresholds
with the D-instanton corrections to the type I thresholds. 
A similar discussion applies to the threshold corrections to the $F^4$ terms
involving the K-K gauge fields. The analysis for the 
type I dual of the type II (4,0) model may be carried over 
to the present situation after taking into account
the presence of a non-vanishing contribution not only 
from the Klein-bottle amplitude ${\cal K}$ but also from the annulus
${\cal A}$ and the M\"obius-strip ${\cal M}$ amplitudes. 
All these surfaces allow
for the closed-string insertions that are needed to extract the thresholds.
A closely related discussion applies to the ${\cal R}^4$ terms.
The technical details needed to clarify the above issues are under 
investigation. 

\vskip 0.5in
{\bf Acknowledgements}

This research has been supported in part by EEC under the TMR contracts
ERBFMRX-CT96-0090 and FMRX-CT96-0012.
 
\rnc{\Large}{\normalsize}

\end{document}